\documentclass[useAMS,usenatbib,a4paper]{mnras}
 \usepackage{epsfig}
  \usepackage{relsize}
  \usepackage{josty}
 \usepackage{graphicx}\usepackage[usenames,dvipsnames]{color}

\title[Satellite Quenching, Inner Density and the Halo] 
{Satellite Quenching, Galaxy Inner Density and the Halo Environment}

\author[Woo et al.] 
    {\parbox{\textwidth}{Joanna Woo$^{1}$\thanks{joanna.woo@phys.ethz.ch}, 
    C. M. Carollo$^{1}$, 
    S. M. Faber$^{2}$, 
    Avishai Dekel$^{3}$,
    Sandro Tacchella$^{1}$
    } 
\vspace{0.4cm}\\
\parbox{\textwidth}{ 
$^{1}$Institute for Astronomy, Department of Physics, ETH Zurich, 
Wolfgang-Pauli-Strasse 27, CH-8093 Zurich, Switzerland\\
  $^{2}$University of California Observatories/Lick Observatory,
  Department of Astronomy and Astrophysics, University of California, 
  Santa Cruz, CA 95064, USA\\
 $^{3}$Center for Astrophysics and Planetary Science, Racah Institute
  of Physics, The Hebrew University, Jerusalem 91904, Israel\\
}}

\pagerange{\pageref{firstpage}--\pageref{lastpage}} \pubyear{2015}

\begin{document}
\label{firstpage}
\maketitle
\begin{abstract}
Using the Sloan Digital Sky Survey, {we adopt the sSFR-$\sigone$ diagram as a diagnostic tool to understand quenching in different environments.}  sSFR is the specific star formation rate, and $\sigone$ is the \jwb{stellar} surface density in the inner kpc.
Although both the host halo mass and group-centric distance affect the satellite population, we find that these can be characterised by a single number, the quenched fraction, such that key features of the sSFR-$\sigone$ diagram vary smoothly with this proxy for the ``environment''.  Particularly, the sSFR of star-forming galaxies decreases smoothly with this quenched fraction, \refc{the sSFR of satellites being 0.1 dex lower than in the field}.  Furthermore, $\sigone$ of the transition galaxies (\ie, the ``green valley'' or GV) decreases smoothly with the environment \refc{by as much as 0.2 dex for $\Ms = 10^{9.75-10}\Msun$} from the field, and decreasing for satellites in larger halos and at smaller radial distances within same-mass halos.   
We interpret this shift as indicating the relative importance of today's field quenching track vs. the cluster quenching track. 
These \jwb{environmental effects} in the sSFR-$\sigone$ diagram are most significant 
in our lowest mass range ($9.75 < \log \Ms/\Msun < 10$).  
One feature that is shared between all environments is that at a given $\Ms$ quenched galaxies have \refc{about 0.2-0.3 dex} higher $\sigone$ than the star-forming population .
\blue{These results indicate that either $\sigone$ increases (subsequent to satellite quenching), or $\sigone$ for individual galaxies remains unchanged, but the original $\Ms$ or the time of quenching is significantly different from those now in the GV}.

\end{abstract}


\section{Introduction}
\label{introduction}

Identifying the physical mechanisms that cause and sustain the cessation of star formation in galaxies has been a persistent problem in galaxy evolution research.  Studies of large surveys over the last decade have made significant progress toward understanding this problem.  In particular, it is now firmly established that the galaxy population obeys three broad correlations \jwb{between morphology, environment and quenching: 1) quenched galaxies tend to have early-type morphology or dense structure, 2) quenched galaxies tend to live in dense or massive halo environments, and 3) galaxies with early-type morphologies are found in these dense environments. }  
\subsection{Quenching and Morphology/Structure}

Ever since Hubble classified the galaxies \citep{hubble26} it has been known that the galaxy population can be roughly divided into star-forming disks and quenched spheroids.  
Early type morphologies or high stellar density strongly correlate with being quenched in the local universe 
(eg., \citealp{str01,kau03,bla03,bell08,vandokkum11,fang13,omand14,schawinski14,bluck14})
and
at high-$z$ (\citealp{wuyts11,cheung12,bell12,szomoru12,wuyts12,barro13,lang14,tacchella15b}).

One manifestation of this correlation between morphology and quenching is that the mass profiles of quenched and star-forming galaxies differ within the inner 1 kpc such that quenched galaxies are denser in this inner region \citep{fang13}.  (This is illustrated in \fig{cartoon} which is a schematic diagram of sSFR vs. the stellar density within 1 kpc $\sigone$.  The quenched population, marked ``Q,'' has at high $\sigone$ relative to the star-forming population, marked ``SF'' .)
At \jwb{any time}, this central stellar density grows along with stellar mass while on the galaxy main sequence (\citealp{hopkins09,feldmann10,vandokkum15}).  The high inner densities seen in quenched galaxies today were already in place by $z\sim 2$ \citep{patel13,barro15,tacchella15b}, which may suggest that if there is a connection between quenching and high inner density, it likely occurs at high-$z$.  

Several ideas have been proposed to explain this observed link between morphology and quenching.  Major dissipative mergers for instance have been demonstrated \refc{in simulations} to build a central spheroid component through violent relaxation of pre-merger stars \citep{toomre72,mihos94,hopkins09}.  The dissipative inflow of gas triggers a central starburst that, along with contributing to the bulge, also quenches a galaxy through fast consumption and/or winds.  \refc{In simulations, the same inflow can accompany disk instabilities \citep{friedli95,immeli04,bournaud11_gasrichmergers} which are observed in local galaxies (\citealp{courteau96,macarthur03,carollo98,carollo01,carollo07}; see also the review by \citealp{kormendy04}.
Violent disk instabilities are predicted in simulations to be especially relevant at high-$z$ \citep{noguchi99,bournaud07,dekel09,mandelker14,dekel14}.  Instabilities can also be triggered by mergers \citep{barnes91,mihos96,hopkins06,zolotov15} and counter-rotating streams \citep{danovich15}}.  
{A candidate merger sequence has been identified at $z\sim 0$ \refc{observations} extending all the way from disturbed starbursts followed by fading, AGN turn-on, and eventual quiescence (\citealp{yesuf14}; see also \citealp{ellison11}).}

While there is a structural correlation with quenching, its causality arrow is not clear.  It may be the effect of quenching that is unrelated to morphology \citep{lillycarollo16}.  It has also been argued and shown that a spheroid can and sometimes does regrow a star-forming disk (e.g., \citealp{brennan15,graham15}) unless accretion is prevented, for example in a hot massive halo \citep{gabor15,woo15,zolotov15} and/or by AGN (\citealp{dekbir06,kormendy13} and references therein).  
In addition, a significant number (25-65\%) of quenched galaxies appear to have (early-type) disks,
especially at high-$z$
(\citealp{stockton04,mcgrath08,vanD08,vandenbergh09,bundy10,oesch10,vanderwel11,salim12,
bruce12,bruce14}), and also a number of bulges appear to be blue 
(e.g., \citealp{carollo07}).  

\subsection{Quenching and the Halo Environment}
\label{intro_quench_env}

The quenched fraction for central galaxies (those that are in the centres of their halo potential wells) correlates with various measures of the halo environment, including the halo mass $\Mh$ (\citealp{woo13,gabor15,woo15,zu16}).  This is often attributed to virial shock heating of infalling gas in haloes more massive than $\Mcrit \sim 10^{12} \Msun$ \citep{croton06,dekbir06}.
Below this critical mass, accretion is cold and conducive to star formation, while above it, 
infalling gas reaches the sound speed and a stable shock forms, heating the gas \citep{ree77,bir03,ker05,dekbir06}.  For a population of haloes,
this is manifested as a mass range extending over one or two decades around $\Mcrit$ where there is a decrease
in the cold accretion as a function of halo mass ($\Mh$)
\citep{ocv08,keres09,vandevoort11}.  

Since the effects of this so-called ``halo quenching'' are expected to increase with the mass of the halo rather than galaxy stellar mass, they are seen most dramatically in the satellite population since their host haloes can be much more massive than those of field galaxies at the same stellar mass.
These effects are seen as ``environment'' quenching in which the galaxy population is more likely to be red or have low sSFR in dense environments
\citep{hog03,kau04,bal04,blanton05b,baldry06,bundy06,coo08,bamford09,patel09,
skibba09a,wilman10,quadri12,haas12,wetzel12,hartley13,knobel15,carollo16} and in massive haloes \citep{woo13,woo15,bluck16}.

{\citet{peng10} argued that this environmental quenching is separable from the quenching that happens in the field (they call this ``mass quenching''), but others} have argued that halo quenching can affect both centrals and satellites \citep{gabor15,woo15,zu16}.  Indeed, the question has been raised as to whether the apparent separability between quenching of centrals and satellites is actually a single phenomenon that masquerades as two separable processes due to the lack of dependence on halo mass of the mass function of the satellite population (\citealp{knobel15,carollo16}), or if quenching is merely a function of halo age (\citealp{hearin13}).
It remains an open question whether satellites experience additional quenching mechanisms than field galaxies, or merely more of the same physics that also quenches field galaxies.

\subsection{Morphology/Structure and the Halo Environment}

The morphology of galaxies is also correlated with dense environments (\citealp{holmberg40,dre80}).  There is evidence that this so-called ``morphology-density'' relation is not due to environmentally-driven morphological change, but only to the increased quenched fraction in dense environments.  For example, the early-type fraction of satellites, {determined by visual classification}, does not rise as steeply with decreasing group-centric distance as the quenched fraction of satellites (\citealp{bamford09}), and the early-type fraction of {\it quenched} satellites {determined by a combination of parametric and non-parametric fitting)} does not vary at all (\citealp{carollo16}) with distance.  These authors use these observations to argue that the morphology-density relation is driven by the increase of the red fraction in dense environments rather than a real increase of early-type morphology in dense environments.  

In fact, if there are two separate quenching channels for satellites as in \citet{peng10},
\cite{carollo16} argue that both channels must have the same effect (either no or some effect) on the morphologies of satellites.  
{If they are the same effect, then it is possible that a single mechanism governs the quenching of galaxies, the same mechanism that operates in the field (which is correlated with morphology/structure), and is simply more advanced in clusters.  This idea has not yet been studied in detail.
}

\subsection{Goals of this Paper}
\label{goals}

\begin{figure}
 \includegraphics[width=0.45\textwidth]{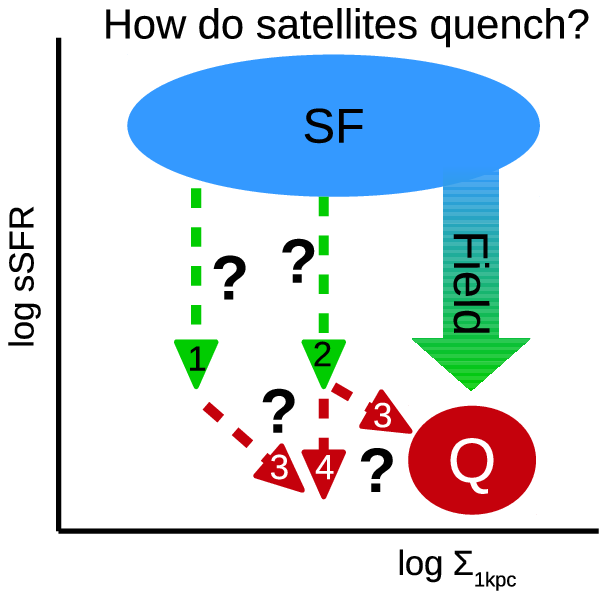}
 \caption{Schematic diagram illustrating quenching paths in sSFR-$\sigone$ space at constant stellar mass.  The region marked ``SF'' refers to the ``star-forming'' population, and the region marked ``Q'' refers to the ``quenched'' population.  The possible satellite quenching paths are the subject of this study. The numbered arrows are discussed in \sec{goals}. }
\label{cartoon}
 \end{figure}

In this paper we aim to study the interplay of the above three correlations for satellite galaxies in order to clarify the role of the halo environment versus the role of morphology/structure-dependent quenching in governing their evolution.  The diagnostic we will use is the diagram of the specific star-formation rate (sSFR) versus the stellar surface mass density within 1 kpc ($\sigone$).  The distribution of galaxies in this parameter space has been used for understanding structural change during quenching for centrals and their quenching track (labelled ``Field'' in \fig{cartoon} - see \citealp{fang13,barro13,tacchella15b}).  This diagram has not yet been compared between environments.  

Do satellites in different environments populate different regions of this diagram?  Such a comparison will aid in understanding simultaneously how star-formation activity and galaxy structure evolve for satellites and how this evolution differs from that of isolated galaxies.  
The various satellite quenching mechanisms proposed in the literature - ram pressure stripping, starvation, harassment - are expected to act on satellites regardless of their $\sigone$.  If such mechanisms are important in satellite evolution, their quenching track in sSFR-$\sigone$ space should resemble something like the arrow \refc{labelled ``2'' in \fig{cartoon}, \ie, in a different location than that of the field.  In particular, if galaxies with low $\sigone$ tend to have more loosely bound gas, ram pressure stripping in particular may preferentially affect galaxies with low $\sigone$, resulting in a quenching track more like arrow ``1'' in \fig{cartoon}.   Some of these mechanisms may cause an increase in $\sigone$ via some triggered star formation during or after quenching (arrows ``3''), or nothing more may happen to the satellites (arrow ``4'').  
This paper attempts to detect any of these effects. }

Even if satellites begin to quench independently of their $\sigone$, their subsequent evolution in sSFR-$\sigone$ and even $\Ms$ may not be \jwb{simple to predict}.  Do satellite quenching mechanisms eventually change $\sigone$ or $\Ms$?  The position of the quenched satellite population in this diagram relative to the position of transitioning satellites will shed light on these issues.  

Note that the above correlations in the literature between morphology, environment and quenching were discovered using simply the quenched fraction.  However, higher-order information can be gained about the transition process as sSFR declines by studying the entire distribution of sSFR rather than just the quenched fraction (as emphasised by \citealp{woo15}).  This kind of study is required for answering the questions posed here.

The literature is replete with many different measures of ``structure'' and ``morphology''.  However, since they all strongly correlate with each other, the above correlations are likely to be true in broad strokes, regardless of which measure is used.  In this paper, we adopt $\sigone$ (introduced by \citealp{cheung12} and \citealp{fang13}) as our fiducial measure of structure for a number of reasons:  
\begin{enumerate}
 \item Despite some uncertainties in its measurement (such as the effect of the PSF), it does not depend on galaxy radius, which is usually light-based in current catalogues (unlike, for example, the density within the effective light radius, $\Sigma_{e}$.)
\item It is measured directly from the data and not a fit to a model profile, which can be plagued by degeneracies and result in large systematic residuals.  
\item It refers to the innermost part of a galaxy, which
is arguably less affected by {stripping}.  
\item \jwb{The galaxy population seems to obey the same $\sigone$-$\Ms$ relation at low- and high-$z$ (\citealp{tacchella15b,tacchella16a} - or with only slight variation \citealp{barro15}), suggesting that $\sigone$ is a stable quantity that on the whole either stays constant with time or increases with time (as stellar mass increases). For individual galaxies, it is difficult to imagine how it could decrease.} 
\end{enumerate}

{Clearly,} 1 kpc means very different things for galaxies of very different sizes (and thus masses).  For this reason, \jwb{we also confirmed that the same analysis using the bulge mass \citep{mendel14}  the \citet{sersic68} index $n$ \citep{simard11} instead of $\sigone$ produced similar results.}  This is after all, not unexpected given the strong correlations between different measures of ``morphology'' and ``structure.''

\sec{data} describes our selection and quantities used from the Sloan Digital Sky Survey (SDSS).  \sec{results} presents our findings regarding the distribution of satellites in the sSFR-structure plane compared to the field and as a function of the halo environment.  \sec{discussion} discusses possible implications of our results on the nature of satellite quenching.

This analysis assumes concordance cosmology: $H_o = 70~{\rm km~s^{-1}
  Mpc^{-1}}, \Omega_M = 0.3, \Omega_\Lambda=0.7$.  Our halo mass estimates also assume $\sigma_8=0.9, \Omega_b =0.04$.

\section{Data}

\label{data}

\subsection{The sample}
\label{sdsssample}

The SDSS \citep{york00,gunn06} sample used here is from Data Release
7 \citep{abazajian09}. This sample
contains 53 603 galaxies after matching all catalogues and applying
all cuts as described below.

We limited the redshift range to $0.03 < z < 0.07$ (in order to minimise seeing effects; see below).  
Applying the K-correction utilities of \cite{blanton07} (v4\_2) on $ugriz$ photometry 
(``petro'' values - \citealp{gunn98,doi10}), using redshifts from the NYU-VAGC (DR7) 
catalogue \citep{blanton05,adelman08,padmanabhan08} and the $r$-band limit
(17.77) of the spectroscopic survey, we
calculate $V_{\rm max}$.  Each galaxy is weighted by its
$1/V_{\rm max}$ multiplied by the inverse of its spectroscopic
completeness (also from the NYU-VAGC) in order to compute all quoted galaxy
fractions and volume densities.

The NYU-VAGC contains 2 506 754 objects, of which 153 662 lie within our chosen
redshift range and are primary spectroscopic targets, 140 758 also have $r$-band 
magnitudes $< 17.77$, and 137 756 of which {\tt KCORRECT} produced finite
$V_{\rm max}$.  {This sample was matched with other catalogues as described below.}

\subsection{Galaxy Central Density}
\label{morph}

Throughout the analysis, we make use of the stellar surface density within the inner
kiloparsec $\sigone$ in order to probe quenching mechanisms which result in centrally
concentrated galaxies.  
This is computed in the following way.  Using the SDSS DR7
Catalogue Archive Server, we retrieved surface
brightness profiles in the $ugriz$ band-passes and corrected them for galactic extinction
using the extinction tags in the {\tt SpecPhoto} table.  The
flux within each radial bin in all five bands is used as input into the
K-correction utilities of \cite{blanton07} (v4\_2) to compute the
stellar mass profile.  These bins are summed to produce the 
cumulative mass profile.  Interpolating between the radial
bins, we estimated the total mass and surface density within 1 kpc. 
(We confirmed that the largest radial bin for our final sample is greater
than 1 kpc, and that their smallest radial bin is smaller than 1 kpc; \ie, it was not necessary 
to extrapolate to 1 kpc for any of our sample.)

This computation does not account for contamination from near neighbours or from 
effects of atmospheric seeing.  However, \cite{woo15} showed that neither of these
are important if the sample is selected such that the PSF FWHM 
(of a double Gaussian fit, \ie, the {\tt psfwidth} in the $r$-band) is $< 2$ kpc.  

Out of 859 281 galaxies with profiles retrieved from CasJobs with $z <
0.3$, 142 852 make the redshift cut.  We further selected
142 363 objects which are spectroscopically classified as galaxies
({\tt SpecClass}=2 from CasJobs), 135 300 of which also make the PSF cut.

We also cut the sample to exclude those that are highly inclined.  Our cut of 
$b/a > 0.5$ leaves a sample of 97 603.  Since inclined galaxies can only be 
disks (and not spheroids), this cut preferentially removes slightly less massive
galaxies.  These also tend to be green (mid-range $g-r$ colour) since their
inclination reddens a presumably star-forming galaxy.  Since these are disks,
they also tend to have lower $\sigone$.
Matching with the NYU-VAGC (after the above cuts) then  produces 90 681 objects.

\subsection{Stellar Masses and Star Formation Rates}
\label{mssdss}

Stellar mass and SFR estimates are provided
online by J. Brinchmann et
al.\footnote{http://www.mpa-garching.mpg.de/SDSS/DR7/\label{brinchsite}}.  These are
derived through SED fitting of spectral features.
The formal $1$-$\sigma$ errors (from the 95$\%$ confidence intervals of the probability\
density distribution) are typically $\ltsima 0.05$ dex.  

SFR estimates are an updated version of those derived in \cite{bri04}.
The \cite{kro01} IMF is used to calibrate these SFR estimates.  
Following the method of \cite{bri04}, these estimates combine
measurements from inside and outside the fibre.  Inside the fibre, 
$\Ha$ and $\Hb$ lines are used to compute SFR for star-forming galaxies. 
For galaxies with weak lines or showing evidence for AGN, SFR is estimated
from an empirical relation between the D4000 break and $\Ha$ and $\Hb$ lines
observed in star forming galaxies.  Outside the fibre, Brinchmann et al. 
fit SEDs to the observed photometry 
The means of the probability density distributions of the SFRs inside and outside
the fiber were added resulting in an estimate of total SFR.  

J. Brinchmann (private communication) estimates the typical error of
these SFRs to be about 0.4 dex for star-forming galaxies.  
For quenched galaxies, the errors are 0.7 dex or more, and these SFRs can be considered to be
upper limits rather than measurements.  

This catalogue contains 927 552, of which 
163 686 pass our redshift cut.  In addition, we limit our sample to those with 
$\Ms > 10^{9.75} \Msun$ (102 998) and those for which the mode and mean of the probability density
distribution of the SFR differed by no more than a factor of 20 (removing 2 galaxies). 
Applying the previous cuts reduces the sample to 60 621.

\begin{figure}
\includegraphics{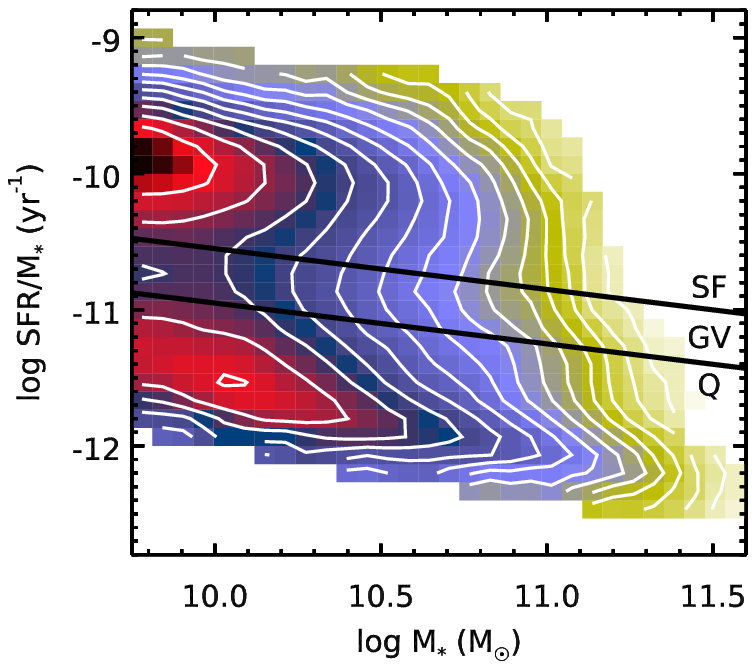}
\caption{\small sSFR vs. $\Ms$.  The solid lines define our divisions between the broad star-forming classes ``SF'', ``GV'' and ``Q'' (used in Figs. \ref{histograms}-\ref{ssfr} and \ref{trendswithqf}).} 
\label{gmsdefs}
\end{figure}

\refc{This study refers to ``star-forming'' (SF) galaxies, those in the ``green valley'' (GV) and those that are ``quenched'' (Q).  Since the main environmental effects in sSFR-$\sigone$ distribution are visible to the eye, they do not depend on any formal definition of these three populations, so we often refer to them qualitatively.  In order to confirm our qualitative assessments from the sSFR-$\sigone$ diagrams, we investigate these three populations individually, and in these cases, we use the formal definition that is shown in \fig{gmsdefs}.  The lines were determined by eye and are defined as $\log {\rm sSFR (yr^{-1})} = 0.2 - 0.92(\log \Ms/\Msun - 10.) \pm 0.2$.
Our results are not sensitive to reasonable variations in the slopes and zero-points of these lines.
}

\subsection{The Group Catalogue and Halo Masses}
\label{groupcat}

\cite{yang12} constructed a group catalogue and estimated group halo
masses for the SDSS DR7 sample based on the analysis of \cite{yang07}.
We use this catalogue with a slight modification as described in 
\cite{woo15} and which we briefly describe here.

\cite{yang07} applied a group finder that estimates the
number density contrast of dark matter particles based on the centres,
sizes and velocity dispersion of group members, assuming a spherical
NFW profile \citep{NFW97}.  Testing this group finder on a 
mock catalogue 
\refc{constructed from the conditional luminosity function model (see
\citealp{yang04})},
they found that their algorithm successfully selected
more than $90\%$ of true haloes more massive than $10^{12}\Msun$.

Once the group catalogue was constructed, \cite{yang07} estimated halo
masses by rank-ordering the
groups by group stellar mass and assigning halo masses according to a
halo mass function.  Comparing these halo masses with groups found in
a mock catalogue, they find an rms scatter of about 0.3 dex.

The stellar masses used by \cite{yang07,yang12} are computed using the
relations given by \cite{bel03}, which can underestimate $\Ms$ for
dusty, star-forming galaxies.  
Thus, using the group catalogue of \cite{yang12}, we recompute
group masses using the stellar masses described in \sec{mssdss}, which
are estimated via SED fitting that incorporates dust.  We applied the
same correction for missing members as in \cite{yang07}, 
and used the same technique for
identifying the groups that are complete to certain redshifts.
(Refer to their paper for details.)  We used the
\cite{tinker08} mass function using the \cite{eisenstein98} transfer
function to assign halo masses to the rank-ordered group masses.  Our
modifications to their method produce 
estimates for halo mass that are consistent with their estimates above $\Mh
= 10^{12.6}\Msun$.  In the range $12 < \log \Mh/\Msun < 12.6$, our
estimates are on average 0.03-0.05 dex higher than theirs, which is
expected given that dusty star-formers are found in this range.
This catalogue from \cite{yang12} contains 633 310 galaxies, of which 
587 575 match the stellar mass catalogue in \sec{mssdss}.  

This study compares galaxies which are satellites in groups with 
field galaxies.  Of the 587 575 galaxies with modified $\Mh$, 462 067
are the most massive member of their group.  \refc{415 750 of these are 
in groups of only one member, and these we called the ``field''.}
Satellites are defined in the same way as in \cite{woo15}
and the definition of the dynamical state of the group is taken from \cite{carollo13b}.
In particular, groups 
(with more than one member: 46 317) are assumed
to be relaxed if the most massive member is
also the nearest to the mass weighted centre of the group (true of 37 921 groups)
\citep{cibinel13,carollo13b}.  Otherwise, the group is potentially unrelaxed (8396).   
The first criterion for a satellite is that it must be a member of a relaxed group, and not be the most massive member (a total of 63 757 satellites).  
Second, if it is in potentially unrelaxed groups, a satellite must 
rank as third massive in the group or lower (adding 53 373 satellites).  
The idea behind the second criterion is that the group
halo is most likely associated with the two most massive members.
However if we restricted our sample only to the first criterion, our
results, though much noisier, would remain qualitatively unchanged. 
\tab{yangtable} summarises how the group catalogue is divided into these categories. 

Removing the most massive two galaxies of potentially unrelaxed groups 
clearly removes galaxies that are more massive than the general population.
However, since our analysis compares satellites and the field at the same mass,
this will not change our overall results.

Applying the previous cuts (in redshift, mass, etc), the field sample size reduces to 35 876
while the satellites sample reduces to 17 727.

\begin{table}
\begin{center}
 \caption{\small Breakdown of the Group Catalogue \label{yangtable}}
 \setlength{\tabcolsep}{4pt}
 \begin{scriptsize}
 \begin{tabular}{ c | cc | c | cc }
 \hline\hline
\mcol{3}{l}{Original Catalogue:} & \mcol{3}{c}{633 310} \\
\mcol{3}{l}{After matching \sec{mssdss}:} & \mcol{3}{c}{587 575} \\
 \hline
 \mcol{3}{c|}{Centrals: 462 067} & \mcol{3}{c}{Satellites: 125 508} \\ 
 \hline
 Field: & \mcol{2}{c|}{Group:} & Relaxed:  & \mcol{2}{c}{Unrelaxed:} \\
 415 750  & \mcol{2}{c|}{46 317} & 63 757  & \mcol{2}{c}{61 751} \\
 \cline{2-3}\cline{5-6}
      & Relaxed: & Unrelaxed: &    & Rank = 2: & Rank $>$ 2: \\
       & 37 921 & 8396 &  &  8378  & 53 373 \\
 \hline
\end{tabular}
 \end{scriptsize}
 \end{center}
 \end{table}

We define the relative group-centric distance $\Dist$ of satellites as the ratio of
the projected distance $\Rproj$ of each satellite to the mass-weighted
group centre and the virial radius $\Rvir = 120
(\Mh/10^{11}\Msun)^{1/3} {\rm kpc}$ (\eg, \citealp{dekbir06}), \ie, $\Dist = \Rproj/\Rvir$.

\subsection{Mass Completeness}

\begin{figure}
 \includegraphics[width=0.5\textwidth]{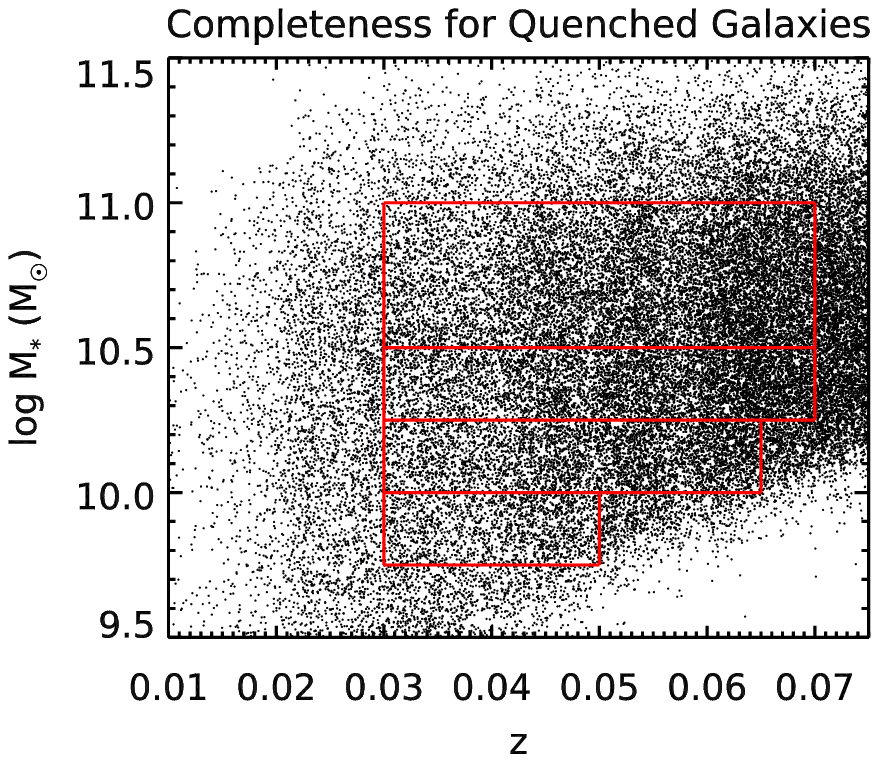}
 \caption{$\Ms$ vs. $z$ for quenched galaxies (as defined in \fig{gmsdefs}).  The red boxes indicated the redshift cuts for the each mass bin that are used in the computation of the quenched fraction (used in \fig{trendswithqf}). }
 \label{completeness}
\end{figure}

We have used the above sample of 53 603 for the majority of our analysis, in particular the sSFR-$\sigone$ diagrams.  One part of our analysis also involves the quenched fraction which will be biased low due to lower completeness for red galaxies.  Therefore in the analysis that compares the quenched fraction between mass bins (\ie, \fig{trendswithqf}), we used more stringent redshift cuts that result in a complete sample for red galaxies.  These cuts are shown in \fig{completeness}.  The resulting samples are 28 894 in the field and 14 820 satellites.

\section{Results}
\label{results}

\subsection{Structural Properties of Satellites and Isolated Galaxies}
\label{structcompare}

\begin{figure*}
\includegraphics[width=0.7\textwidth]{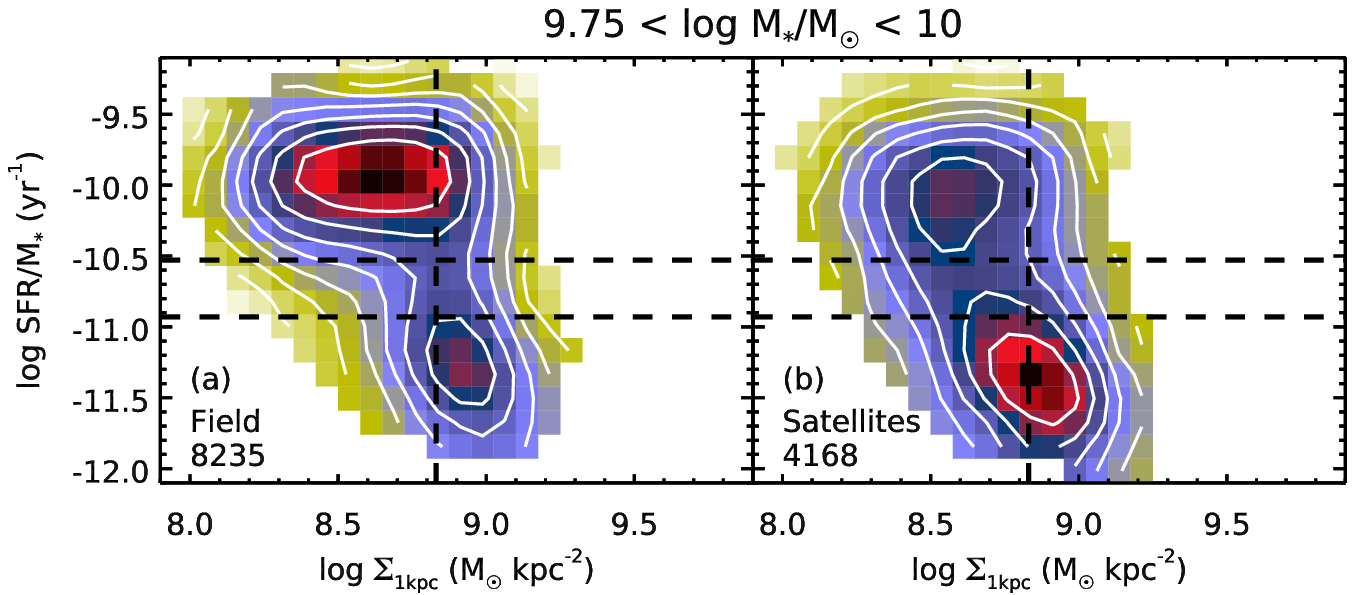}
\includegraphics[width=0.7\textwidth]{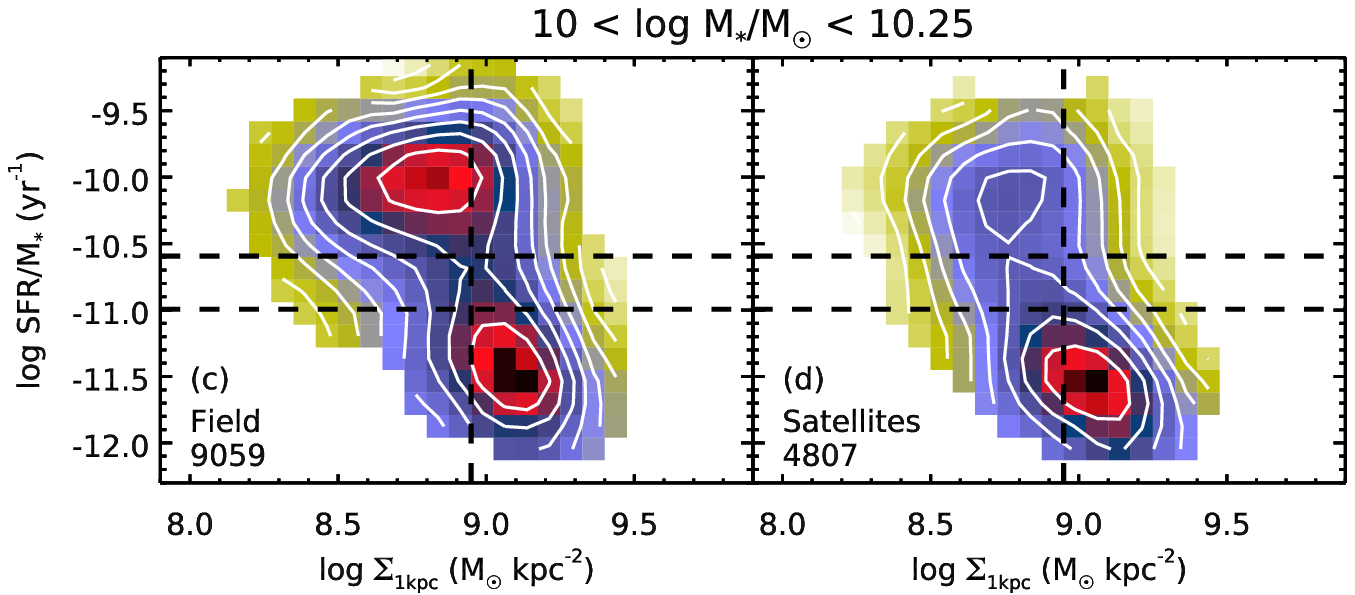}
\includegraphics[width=0.7\textwidth]{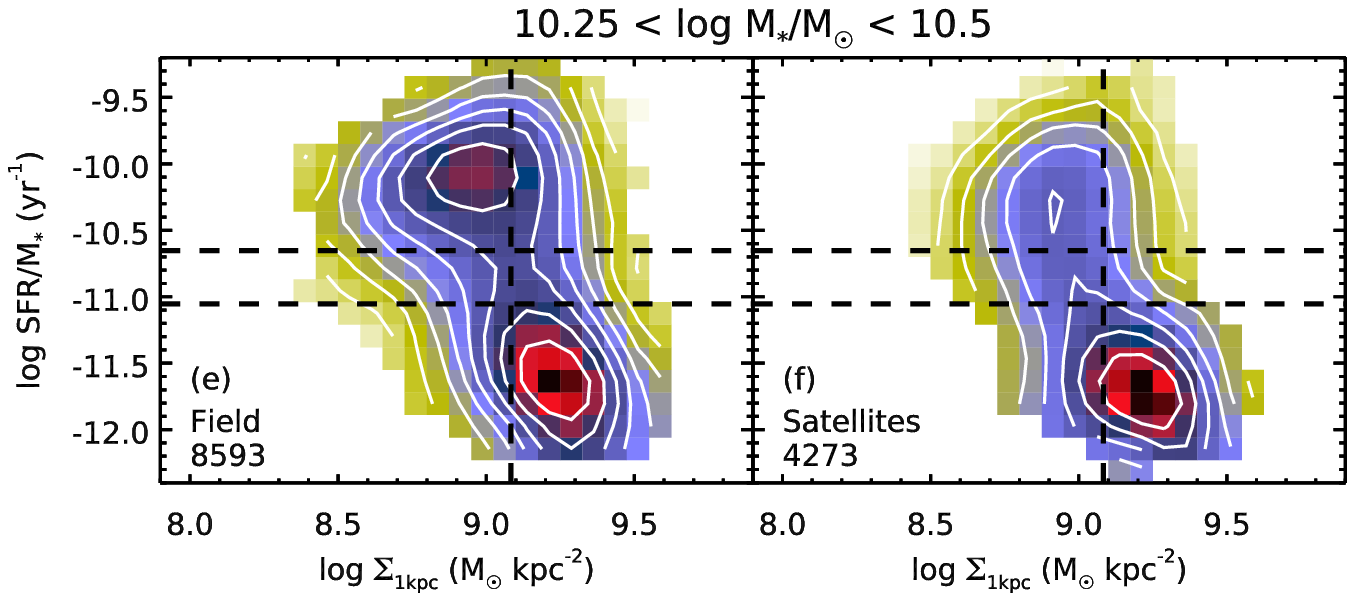}
\includegraphics[width=0.7\textwidth]{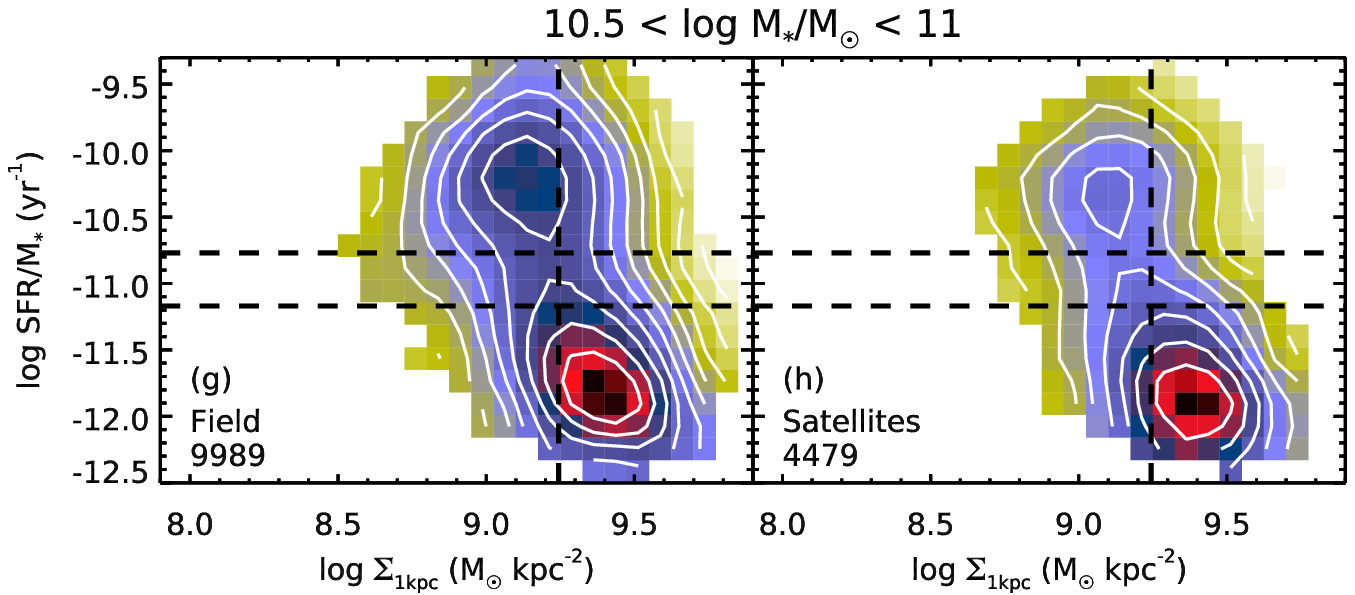}
\caption{\small sSFR vs. $\sigone$ comparing the field to satellites in four $\Ms$ bins.  
The GV as defined in \fig{gmsdefs} lies roughly between the two horizontal lines.  
For reference, the vertical line marks the median GV of the field.
  The contours represent the logarithmic number density 
  in each panel and are separated by 0.2 dex.  The colour scale is normalised such that dark red represents the highest number density in the panel.
  The sample size is indicated in the bottom left.  
  Satellites have a higher proportion of quenched galaxies than the field at all masses.
  The GV bridge connecting the SF and Q populations lies at lower $\sigone$ for satellites than for the field \refc{(by $\sim$ 0.2 dex for the lowest mass bin)}.  
  Quenched galaxies have \refc{higher $\sigone$ than SF galaxies by about 0.2-0.3 dex} whether in the field or as satellites.
  These effects are strongest for the lowest masses, and decrease in significance for the most massive galaxies.
  }
  \label{ssfrvssigone}
\end{figure*}

In order to understand the effect of the halo environment on structural/morphological properties, we compare these properties between satellite and field galaxies.  While we compare these populations in the same $\Ms$ bins, they lie in very different haloes.  An analysis of the effect of the halo environment will follow in the next subsection.

In \fig{ssfrvssigone} we show the distribution of sSFR (SFR/$\Ms$) and $\sigone$ for field (left panels) and satellite galaxies (right panels) in four bins of $\Ms$ from top to bottom.  
The contours represent the logarithmic number density of the population in each panel (weighted by volume and smoothed - the raw points are shown in the Appendix in \fig{ssfrvssigone_dots}.)
The 
contours 
in each panel clearly show the well-known bimodality in star formation between star-forming (SF) galaxies with high sSFR and quenched (Q) galaxies at low sSFR, bridged by a population of intermediate sSFR often called the green valley (GV).  (We formally define these star-forming categories in \fig{gmsdefs}.  These divisions roughly correspond to the two horizontal lines in \fig{ssfrvssigone}.)  
\Fig{ssfrvssigone} shows that for both satellites and the field, in order to be quenched, galaxies need to have relatively high $\sigone$ for their mass bin (\refc{by 0.2-0.3 dex}), as noted by \citet{cheung12} and \citet{fang13}, as well as other authors using different measures of structure/morphology (see \sec{introduction}).  \refc{In other words, there are no quenched galaxies with low $\sigone$ in a given mass bin.}  
  
\begin{figure*}
 \includegraphics[width=0.48\textwidth]{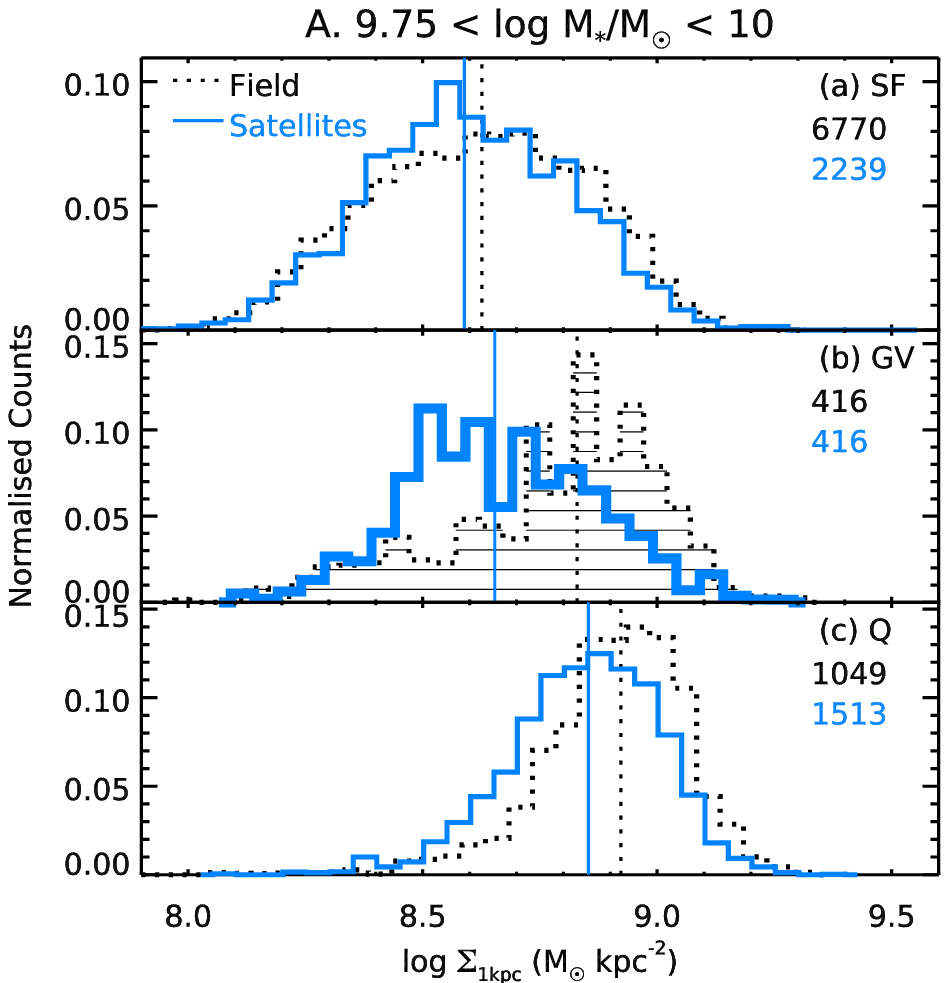}
 \includegraphics[width=0.48\textwidth]{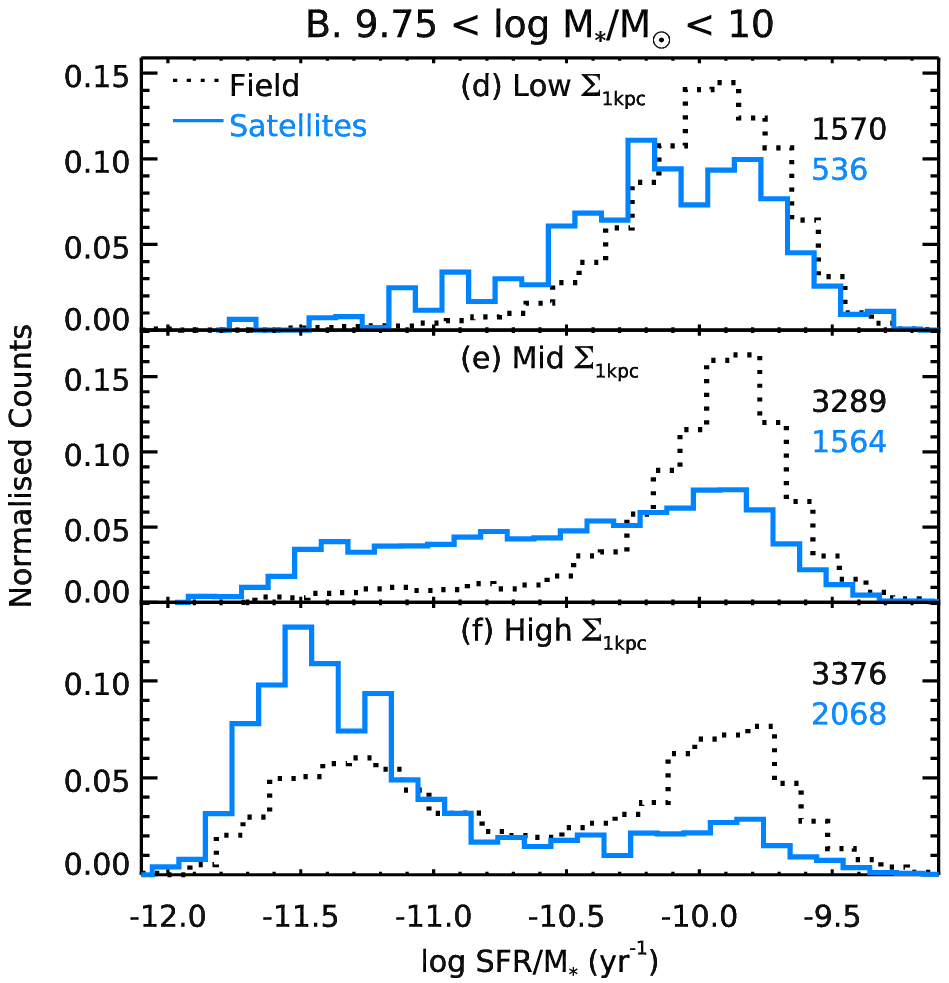}
 \includegraphics[width=0.48\textwidth]{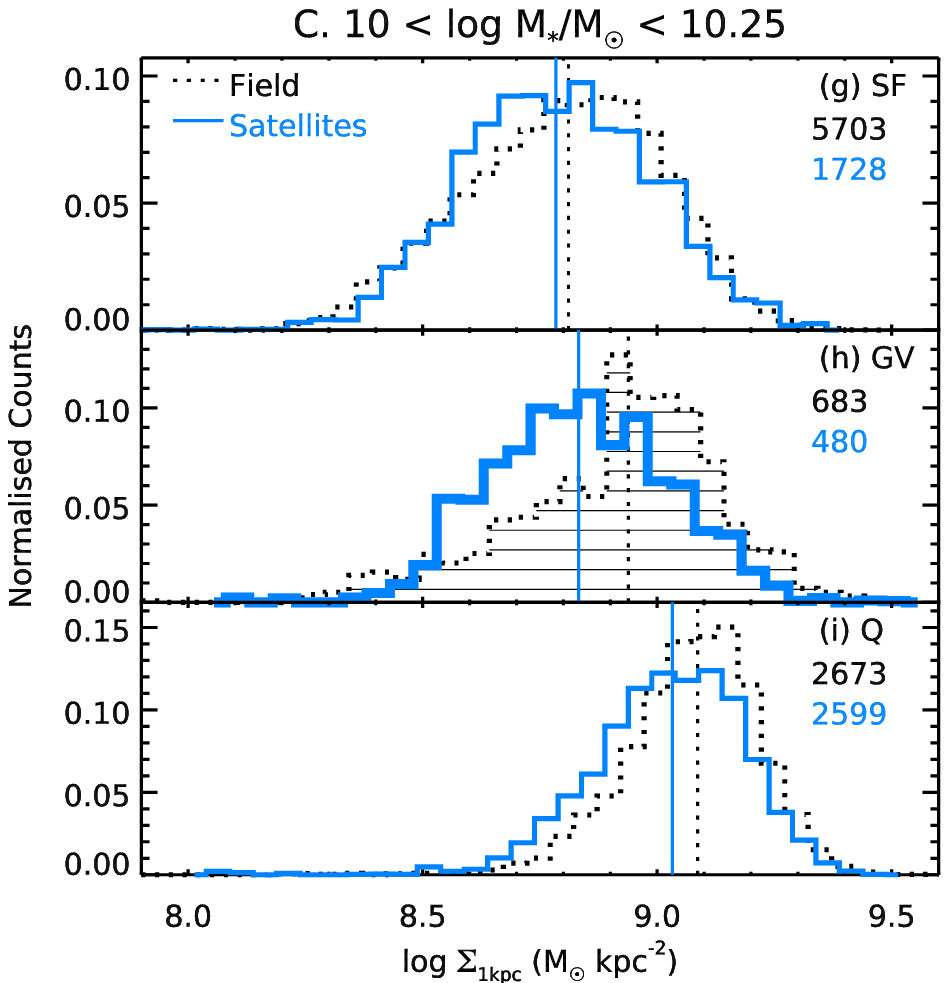}
 \includegraphics[width=0.48\textwidth]{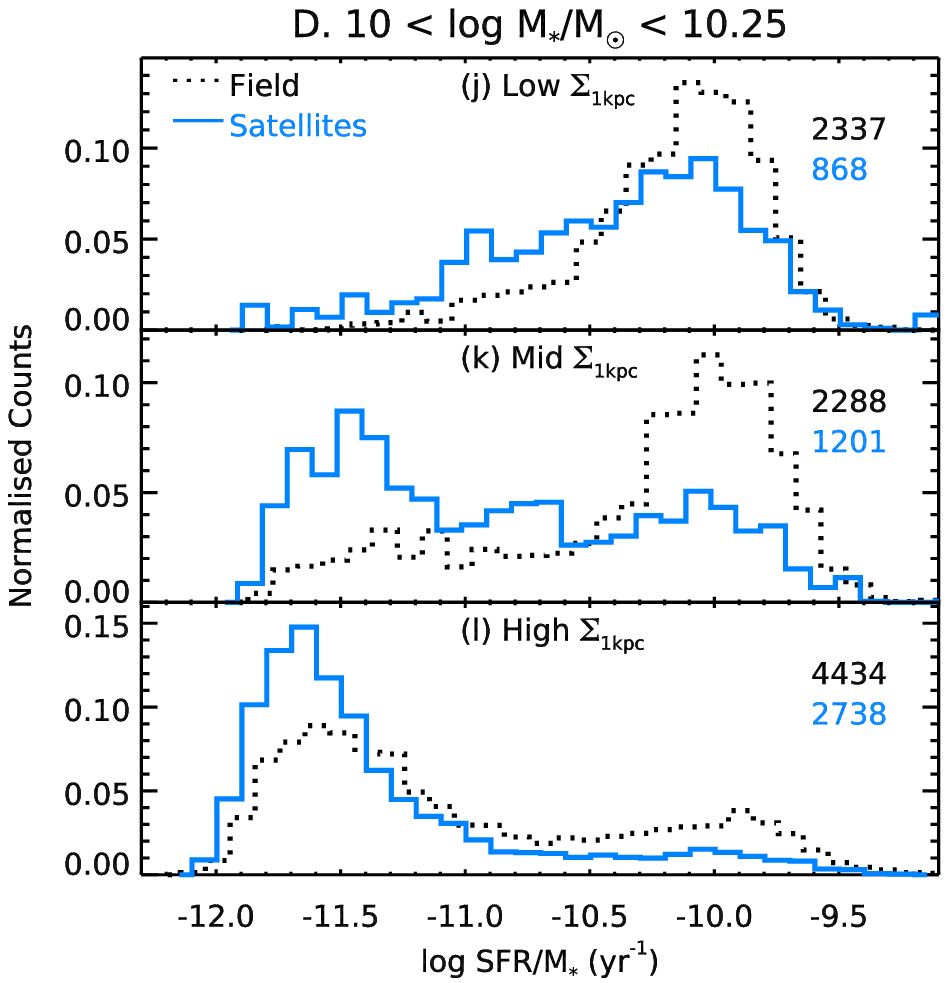}
 \caption{Left column: One-dimensional distributions of $\sigone$ in the mass range $9.25 < \log \Ms/\Msun < 10$ (A) and $10 < \log \Ms/\Msun < 10.25$ (C), in three star-forming classes corresponding to the divisions in \fig{gmsdefs}.  Right column: Distributions of sSFR in three bins of $\sigone$ in the same mass ranges.  The divisions between ``Low'', ``Mid'' and ``High'' $\sigone$ are log $\sigone/(\Msun {\rm kpc}^{-2})$ = 8.4 and 8.7 for the lower mass bin, and log $\sigone/(\Msun {\rm kpc}^{-2})$ = 8.7 and 8.9 for the more massive bin.  The vertical lines in the left panels show the medians of each histogram.  The histograms are normalised by number density.  These histograms confirm these features seen in \fig{ssfrvssigone}: GV satellites lie at lower $\sigone$ than GV field galaxies \refc{by about 0.2 dex for the lowest mass bin} ($b$ and $h$); Q satellites have a lower range $\sigone$ than the field ($c$ and $i$); the peak sSFR of SF and Q satellites is lower than that of SF field galaxies ($d$-$f$ and $j$-$l$). }
 \label{histograms}
\end{figure*}

\begin{figure*}
\includegraphics[width=0.8\textwidth]{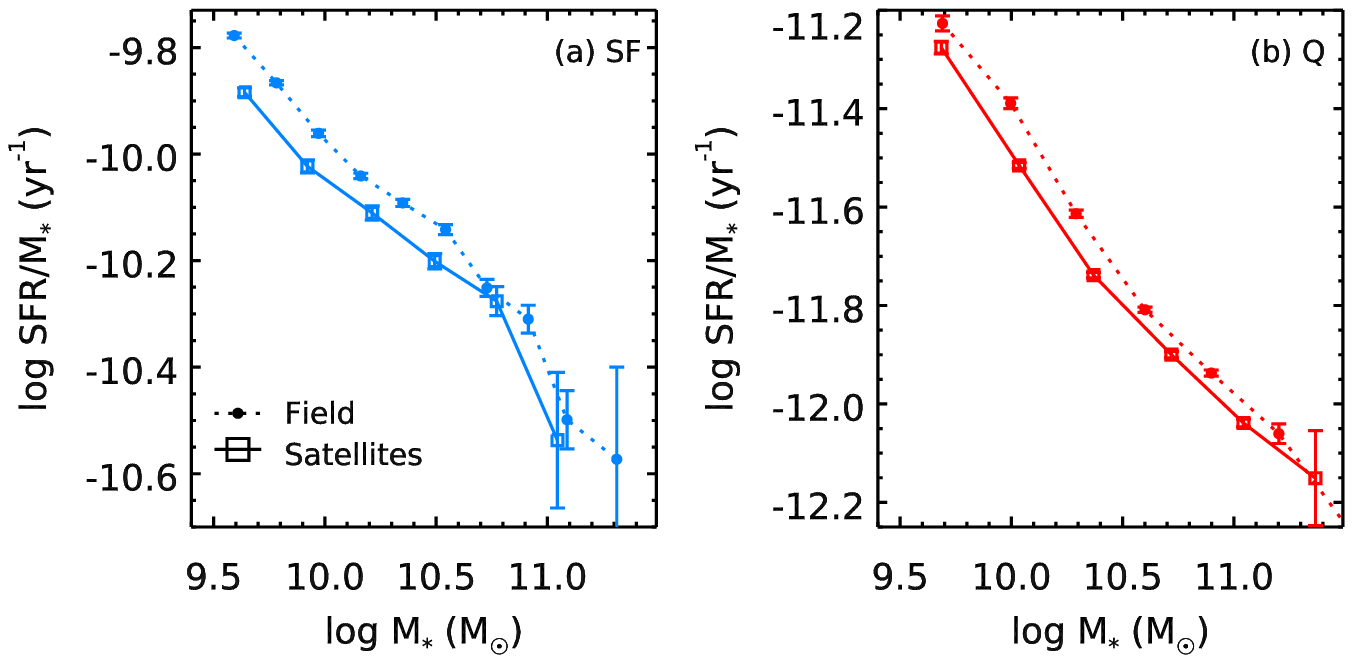}
\caption{\small The median sSFR as a function of $\Ms$ for satellites (square symbols connected by solid lines) and field galaxies (circles connected by dotted lines). The two panels include only SF ($a$) and Q ($b$) galaxies as defined in \fig{gmsdefs}. 
Error bars are bootstrap errors.  
Satellites have lower sSFR than field galaxies \refc{by about 0.1 dex} at the same mass and star formation class.  In particular, the main sequence is lower for satellites than for field galaxies.}
\label{ssfr}
\end{figure*}

But despite this overall similarity, 
\fig{ssfrvssigone} also shows that satellites occupy somewhat different regions of the sSFR-$\sigone$ plane than field galaxies, especially when considering the lowest mass bin (top two panels). 
Apart from the relative number of quenched satellites being higher than the quenched field galaxies in all mass bins (as found in many studies - see \sec{introduction}), there are also some notable differences:  
\begin{enumerate}
\item Relative to the field galaxies, {the GV ``bridge'' between the two populations of the sSFR bimodality is shifted \refc{toward lower} $\sigone$ for satellites.}  I.e., this bridge lies at high $\sigone$ for the field galaxies, and at more mid-range values of $\sigone$ for satellites.  \refc{The effect is a shift in $\sigone$ of $\sim$0.2 dex for the lowest mass bin, and less than 0.1 dex for the highest.}  (For reference, the vertical lines in \fig{ssfrvssigone} show the median $\sigone$ of the {\it field} GV in each mass bin.) 
\item Perhaps related to this, the \refc{lowest-$\sigone$ edge} of the Q population for satellites appears stretched \refc{towards lower $\sigone$} relative to the Q field.  
\item The centroids of both the SF and Q populations are slightly lower in sSFR for satellites (\refc{by about 0.1 dex - see \fig{ssfr} which is described below}; it is also seen in the peaks of the one-dimensional sSFR distribution - see \fig{histograms}).
\end{enumerate}
Note that these observations will be missed in simple studies of the quenched fraction rather than the bivariate distribution in the sSFR-$\sigone$ plane.

The lower panels of \fig{ssfrvssigone} show that most of these differences between field and satellite galaxies listed above are present also at higher masses, but these differences are much smaller than for the lowest bin of $\Ms$.  The GV bridge of satellites is shifted towards lower in $\sigone$ relative to the field, and the centroids of the SF and Q populations are shifted to lower sSFR and lower $\sigone$ respectively (except perhaps for the most massive Q satellites), \refc{but these differences are smaller than 0.1 dex}.

\fig{histograms}A and B flatten the information in \fig{ssfrvssigone}$a$ and $b$ into one-dimensional distributions of $\sigone$ in three star-forming classes (panels $a$, $b$ and $c$) and sSFR in three bins of $\sigone$ (panels $d$, $e$ and $f$).  Satellite histograms are in blue while the field histograms are in black.  SF, GV and Q galaxies are defined by the dividing lines in \fig{gmsdefs}.  (Our results are not sensitive to reasonable variations in slope and zero-points of these divisions.)  The histograms are normalised by their total number densities.  \fig{histograms}C and D show the second mass bin ($10 < \log \Ms/\Msun < 10.25$).  For brevity, we show only our two lowest bins ($9.75 < \log \Ms/\Msun < 10$ and $10 < \log \Ms/\Msun < 10.25$), where the effects are largest.  

\fig{histograms} confirms the same features described above.  The GV for satellites (bold blue histogram in panel $b$ and $h$) is peaked at lower $\sigone$ (\refc{by about 0.2 dex for the lowest masses}) than the GV field galaxies (dotted black histogram) which corresponds to the shift of the GV bridge in \fig{ssfrvssigone}$b$ and $d$.  The Q satellites are also weighted toward lower $\sigone$ than the field but not as much as the GV ($\sim$0.1 dex - panels $c$ and $i$ of \fig{histograms}).  The total $\sigone$ distribution for satellites (adding up the solid histograms) is weighted towards higher in $\sigone$ than the total distribution for field galaxies (not shown here).  (This is related to the morphology-density relation \citealp{holmberg40,dre80}.)  This is the combined effect of quenching being correlated with high $\sigone$ and the satellites having higher quenched fraction than the field.  However at fixed sSFR, GV and Q satellites have lower $\sigone$ than GV and Q galaxies in the field.

It should be noted that even within a narrow mass bin, $\sigone$ is correlated with $\Ms$.  Since the mass function of satellites is bottom-heavier than that of the field (\citealp{peng10}), we checked if the $\sigone$ differences between satellites and the field might be due to the mass differences.  We find that the differences in the mass distributions ($\ltsima 0.03$ dex) are too small to account for the differences we see in $\sigone$.  (The slopes of the $\sigone$-$\Ms$ relation for SF and Q galaxies are $\sim 0.9$ and 0.65 respectively - see also \citealp{barro15}.)

The right panels of \fig{histograms} show that at fixed $\sigone$, the star-forming peak of sSFR bimodality is slightly lower for satellites than in the field \refc{by about 0.1 dex}.  In these panels the tail of the SF satellites seems to fill the GV.  In fact, for the field, galaxies with mid-range $\sigone$ that are quenched are relatively rare, whereas quenched satellites are more than half the satellite population with mid-range $\sigone$ (panels $e$ and $k$).  {The quenched peaks, where present, are also lower in sSFR for satellites than the field.}

We also show this lowering of the sSFR explicitly in \fig{ssfr}, which shows the median sSFR as a function of $\Ms$ for satellites and field galaxies.  This figure shows that SF and Q satellites have lower sSFR than field galaxies at the same mass. 
In particular, the star-forming ``main sequence'' of satellites is systematically lower than that of field galaxies by about 0.05-0.1 dex, compared to a spread of 0.2-0.3 dex for the whole main sequence (depending on how the main sequence is defined).  This difference also disappears around $\Ms \sim 10^{10.7}\Msun$ and higher.

\jwb{Note that these findings are features of the sSFR-$\sigone$ plane, but we checked that a similar analysis using other independent measures of morphology/structure (bulge mass and \Sersic $n$) produces simlar results.  }

\subsection{Structural Properties of Satellites Within the Halo Environment}
\label{environment}

\begin{figure*}
 \includegraphics[width=0.8\textwidth]{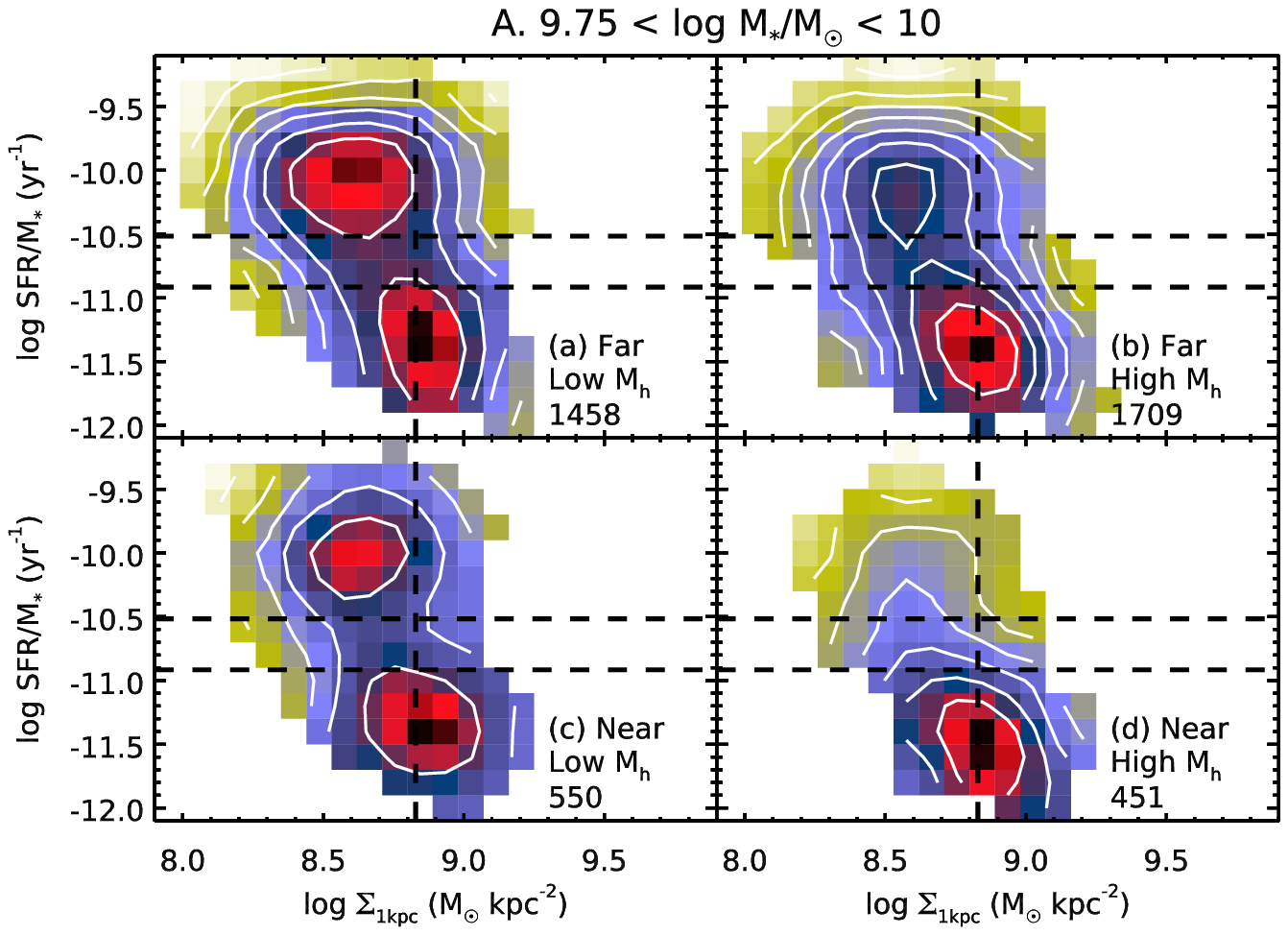}
  \includegraphics[width=0.8\textwidth]{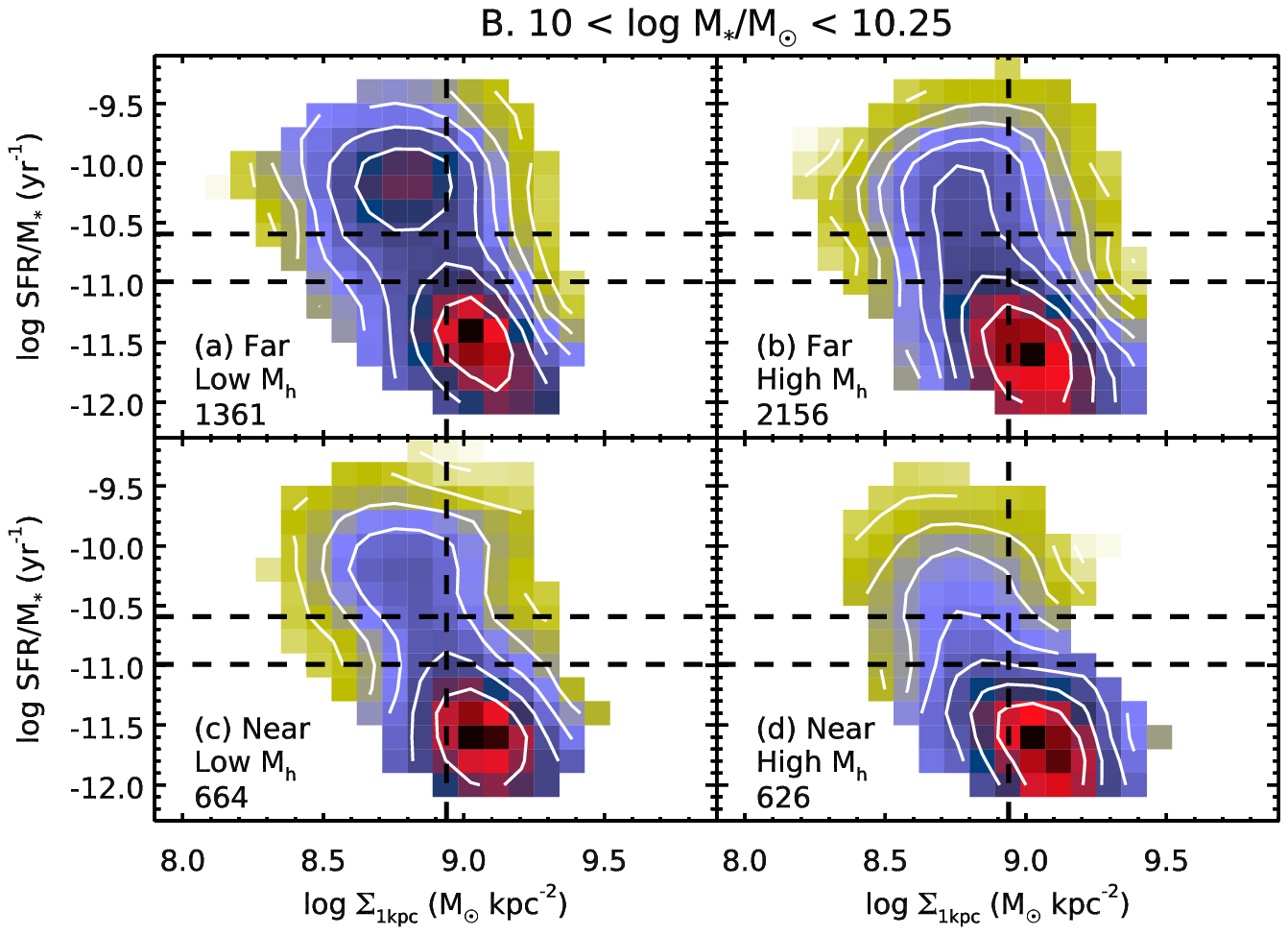}
 \caption{The distribution of satellite galaxies in the sSFR-$\sigone$ plane in four panels representing different environments, for two mass bins (A and B).  ``Near'' and ``Far'' refer to log $\Dist$ less than or greater than -0.6 (or $\Rproj \sim 0.25\Rvir$).  ``Low $\Mh$'' and ``High $\Mh$'' refer to haloes less massive than or more massive than $10^{13.5}\Msun$. 
 Satellites in the GV as defined in \fig{gmsdefs} lie rougly between the two horizontal lines. For reference, the vertical line marks the median $\sigone$ for the field GV.
 The contours represent the logarithmic number density of 
  the population in each panel and are separated by 0.2 dex. The colour scale is normalised such that dark red represents the highest number density in that panel.
  The sample size of each panel is indicated in the bottom corner.  The shift of the GV bridge, the lowering of sSFR for the SF satellites and the extension of Q satellites toward lower $\sigone$ compared with the field is driven by high $\Mh$ and low $\Dist$.
  }
 \label{panelsdvsmh_lowm}
\end{figure*}

\begin{figure*}
 \includegraphics[width=0.8\textwidth]{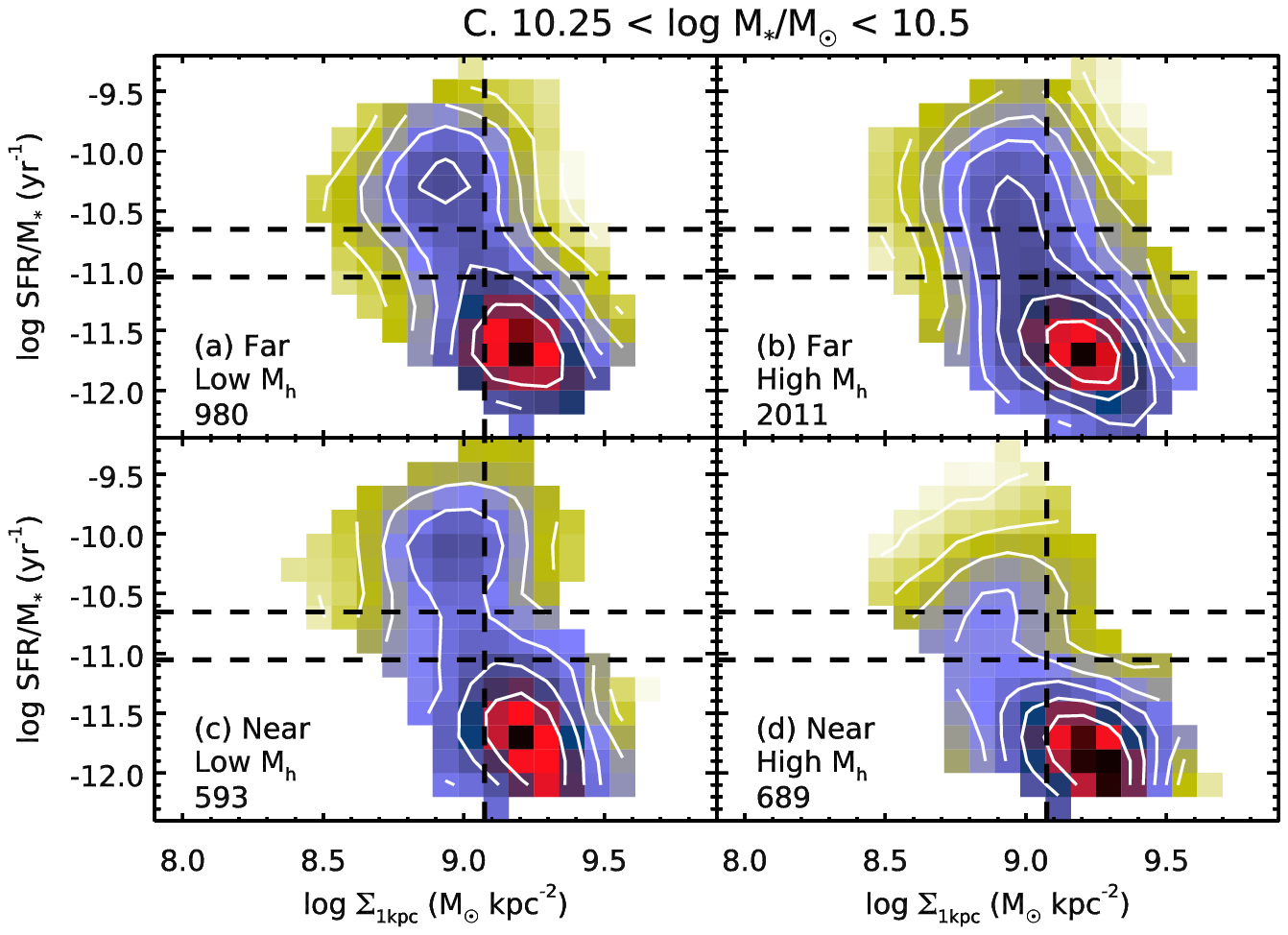}
  \includegraphics[width=0.8\textwidth]{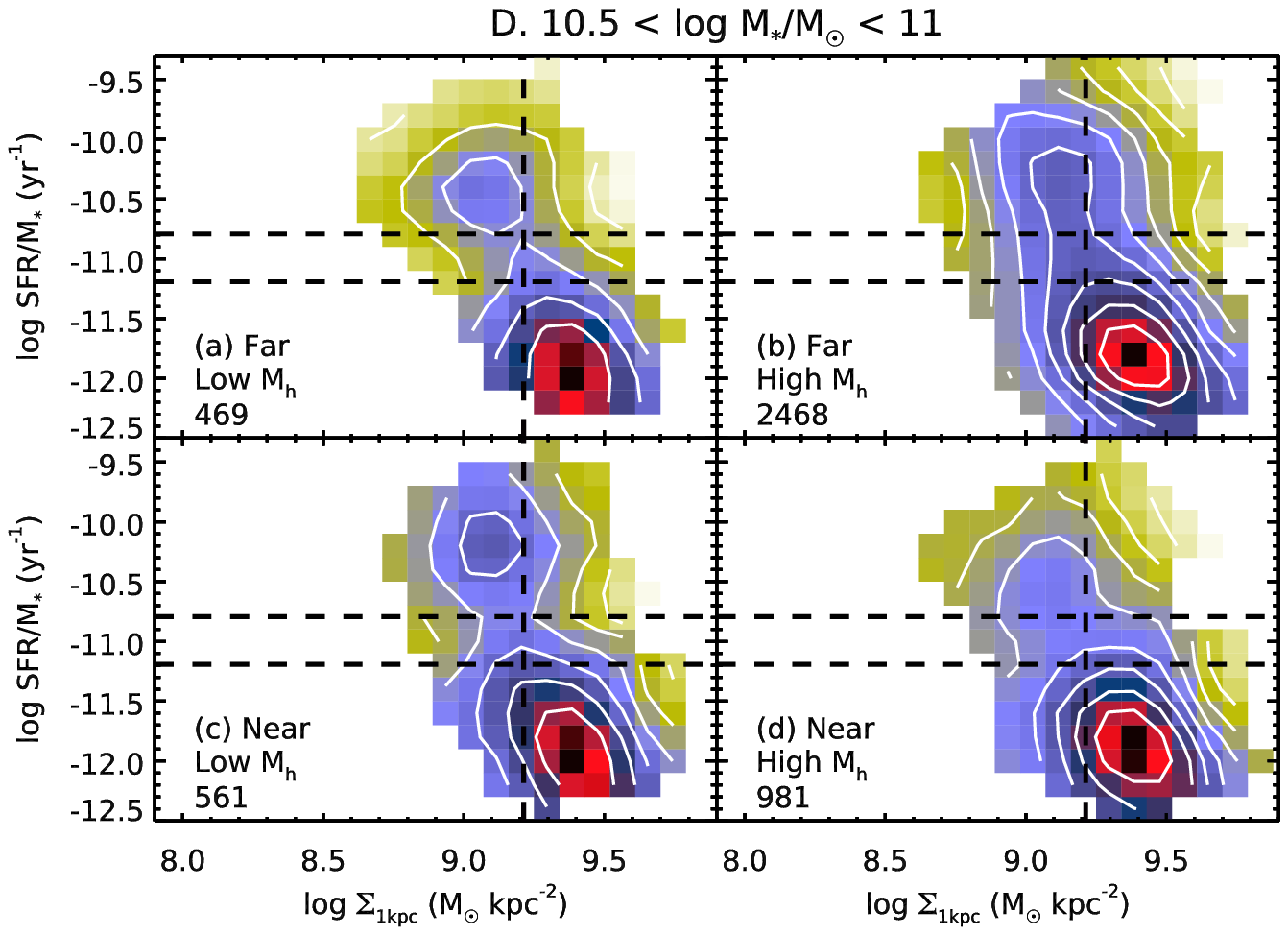}
\caption{The distribution of satellite galaxies in the sSFR-$\sigone$ plane in four panels representing different environments, for two mass bins (C and D).  ``Near'' and ``Far'' refer to log $\Dist$ less than or greater than -0.6 (or $\Rproj \sim 0.25\Rvir$).  ``Low $\Mh$'' and ``High $\Mh$'' refer to haloes less massive than or more massive than $10^{13.5}\Msun$.   Satellites in the GV as defined in \fig{gmsdefs} lie roughly between the two horizontal lines.  For reference the vertical line marks the median $\sigone$ of the field GV.
The contours represent the logarithmic number density of the population in each panel and are separated by 0.2 dex.
The colour scale is normalised such that dark red represents the highest number density in that panel.
  The sample size of each panel is indicated in the bottom left corner.  The massive satellites here continue the trends seen in the two lower-mass bins, but more weakly than the less massive satellites.
   }
 \label{panelsdvsmh_highm}
\end{figure*}

\begin{figure*}
\includegraphics[width=0.95\textwidth]{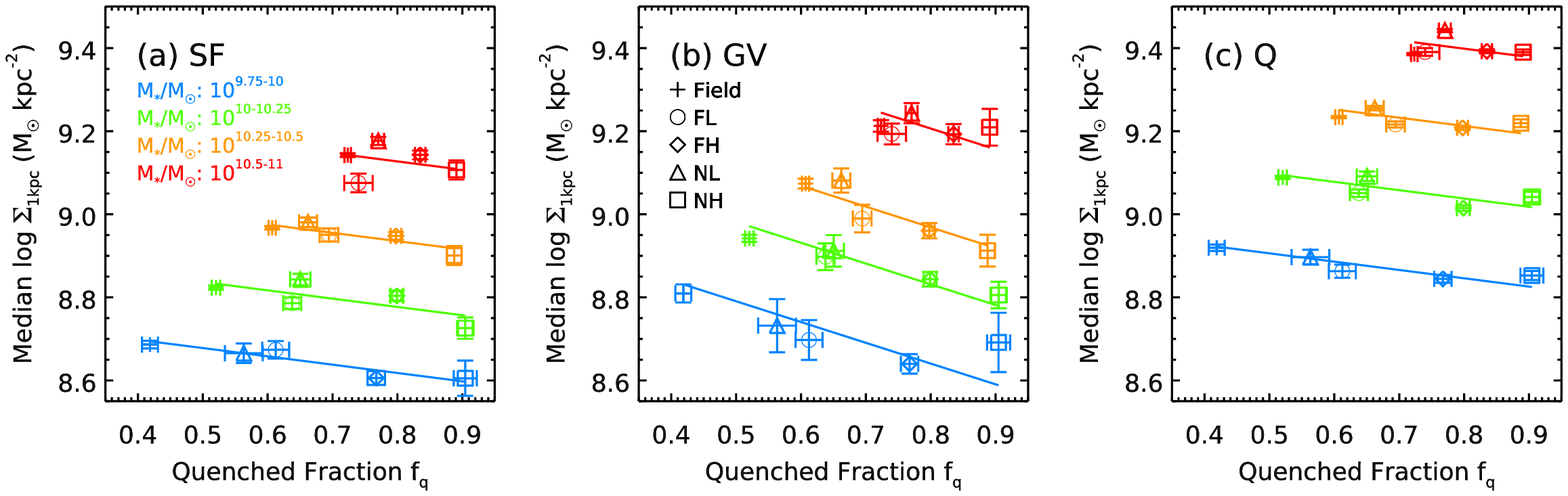}
\includegraphics[width=0.67\textwidth]{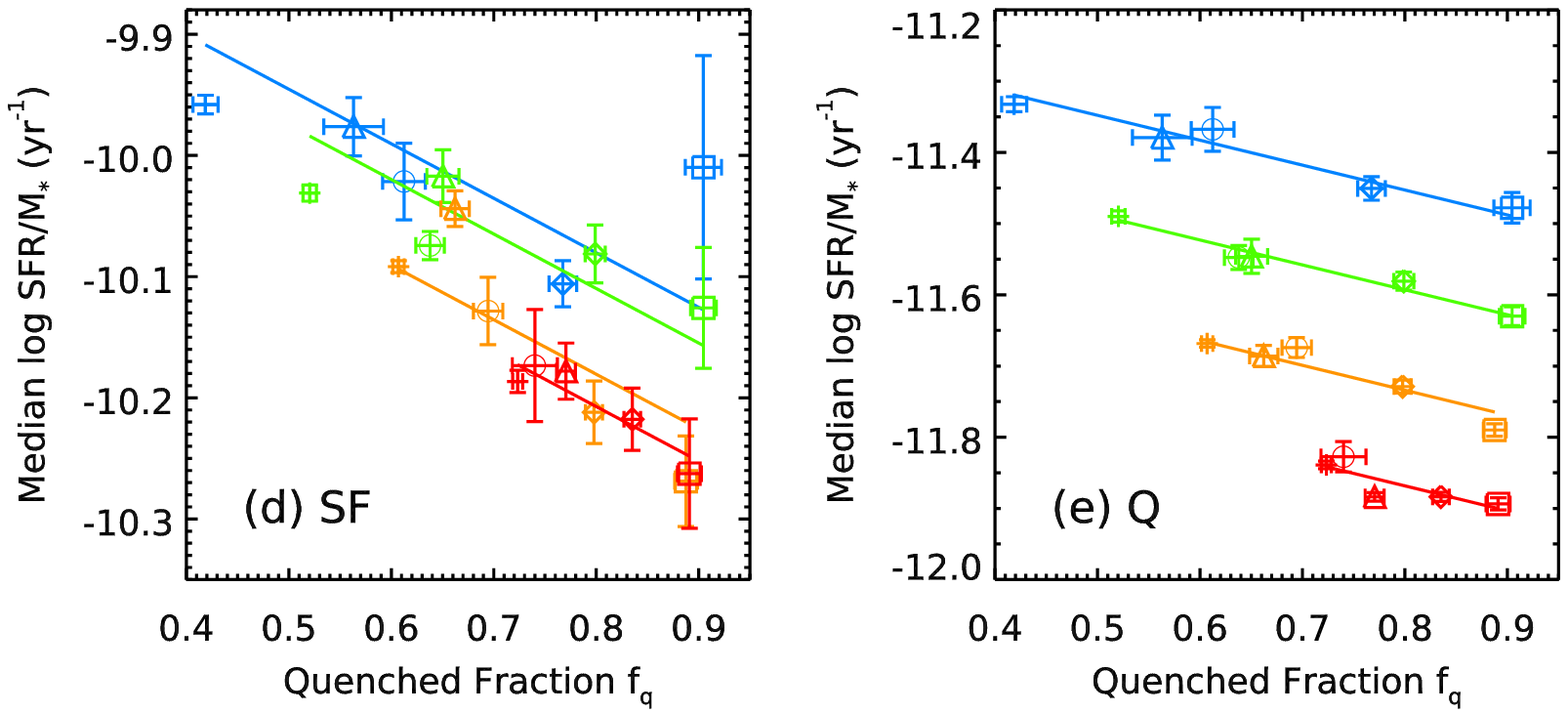}
\caption{$a-c$: The median log $\sigone$ of the SF ($a$), GV ($b$) and Q ($c$) populations in four mass bins as a function of the quenched fraction of different environments (labelled).  
$d, e$: The median sSFR of the SF ($d$) and Q ($e$) populations as a function of $\Ms$ and environment (represented by $\fq$). 
``FL'', ``FH'', ``NL'' and ``NH'' refer to satellites in panels $a$, $b$, $c$ and $d$ of \twofigs{panelsdvsmh_lowm}{panelsdvsmh_highm}.  Error bars represent bootstrap errors.  The lines are fitted to have fixed slopes in each panel: -0.2 ($a$ and $c$), -0.5 ($b$), -0.45 ($d$) and -0.35 ($e$).
The median log $\sigone$ of the GV decreases regularly with quenched fraction, which is our proxy for the environment.  
}
\label{trendswithqf}
\end{figure*}

While the above plots have compared field and satellite galaxies at the same stellar mass, satellites are subject to two other potentially influential parameters, namely, the mass of the halo in which they reside and their distance from the centre of that halo.
Since halo masses are assigned based on the total stellar mass of all members in a group, a satellite must lie in a halo at least twice as massive as a field galaxy of the same $\Ms$.

To disentangle environment, morphology and structure from each other, we divide the $\dist$-$\Mh$ plane into four quadrants at $\Mh=10^{13.5}\Msun$ and log $\dist = -0.6$ ($\Rproj \sim 0.25 \Rvir$), and study the sSFR-$\sigone$ distribution in these different environments. 
\refc{(These divisions were chosen because the quenched fraction depends most stongly on $\dist$ above this mass, and most strongly on $\Mh$ below this distance, as seen in Fig. 10$b$ of \citealp{woo13}. )}
These distributions are shown in different mass bins in \twofigs{panelsdvsmh_lowm}{panelsdvsmh_highm} (Again, the raw unweighted points are shown in the Appendix, \twofigs{panelsdvsmh_lowm_dots}{panelsdvsmh_highm_dots}.)  
In these figures, ``Near'' and ``Far'' refer to satellites with log $\dist$ less than and greater than -0.6, while ``low $\Mh$'' and ``high $\Mh$'' refer to satellites in host haloes less than or greater than $10^{13.5}\Msun$.  
The GV as defined in \fig{gmsdefs} lies roughly between the two horizontal lines, and the vertical line marks the median $\sigone$ of the field GV for reference.

\twofigs{panelsdvsmh_lowm}{panelsdvsmh_highm} show that {for satellites in all environments, low sSFR is associated with high $\sigone$ as is the case for the field} (\fig{ssfrvssigone}).  $\sigone$ for Q galaxies is \refc{always about 0.2-0.3 dex higher than that} of SF galaxies.
The satellites in \fig{panelsdvsmh_lowm}$a$, i.e., those far from their group centres and in less massive haloes, {are the most similar to the field (\fig{ssfrvssigone}$a$).  It is the panel with the lowest quenched fraction, and the SF distribution is wide in $\sigone$, as in the field. }  These 
may only be loosely bound to their groups.  It is 
also quite possible that many of these are actually field galaxies misclassified by the group finder. 
However, despite these uncertainties, a comparison between \fig{panelsdvsmh_lowm}$a$ and \fig{ssfrvssigone}$a$ shows that even this population of satellites is not the same as the field population.  The same phenomena observed in \fig{ssfrvssigone} when comparing the field and the satellites are also present when comparing this subsample of satellites: the GV bridge for satellites is shifted toward lower $\sigone$, the SF and Q populations of satellites are shifted toward lower sSFR, and the spread of $\sigone$ for the Q satellites is elongated toward lower $\sigone$.  

Comparing the other panels of \twofigs{panelsdvsmh_lowm}{panelsdvsmh_highm}, the quenched fraction clearly depends on both $\Mh$ and $\dist$.  This can be thought of as the zero-order effect seen in \cite{woo13} and other studies of the quenched fraction 
\refc{(\citealp{bal04,hansen09,peng10,wetzel12} and others mentioned in \sec{intro_quench_env})}.  
The higher-order effects, \ie, the differences between satellites and the field seen in \fig{ssfrvssigone} (listed in \sec{structcompare}) seem also to depend on these measures of environment.  In fact, the position of the GV seems to shift more towards lower $\sigone$ with more extreme environments (higher $\Mh$ and lower $\dist$) just as the quenched fraction increases.  

Here we explore this observation further.  
We take the quenched fraction $\fq$ in each of the four panels of \fig{panelsdvsmh_lowm}A and B, and put this number on the $x$-axis of \fig{trendswithqf}$b$.  Thus we are treating $\fq$ as a proxy for the ``environment'' even though the enviroment is a function of our two orthogonal measures ($\Mh$ and $\dist$).  The $y$-axis is the median $\sigone$ of the GV population in that panel.  We also do this for the additional ``environment'' of the field, adding the quenched fraction and GV position of \fig{ssfrvssigone}$a$ to \fig{trendswithqf}$b$.  These five environments are represented by the five different symbols, while the colours represent different mass bins.  Panels $a$ and $c$ of \fig{trendswithqf} show the medians of $\sigone$ for the SF and Q populations as well.  (Note that for this figure, we used the mass-complete redshift cuts described in \sec{data} in order to compare quenched fractions.)  \Fig{trendswithqf}$b$ shows that within a mass bin, the median $\sigone$ of the GV declines smoothly with our proxy for the environment, \refc{so that the GV galaxies in the most extreme environment have a $\sigone$ that is $\sim$0.2 dex lower than the field.}  
$\sigone$ of the GV declines with environment at a similar rate in each mass bin. The lines, which have a fixed slope of -0.5, are good fits to each bin.  The median $\sigone$ for SF and Q galaxies also declines somewhat with environment, but this is not as strong as the trend for the GV.  \refc{The maximum difference in $\sigone$ between the enviroments is $\sim$0.1 dex.}  (The lines in panels $a$ and $c$ are fit vertically with a slope of -0.2.)

Panels $d$ and $e$ of \fig{trendswithqf} show the median sSFR of the SF and Q populations as a function of $\fq$, split into the same mass and environment bins.  The lines have a fixed slope of -0.45 and -0.35 for the SF and Q populations respectively.  These panels show that the median sSFR also declines smoothly with our proxy for the environment \refc{so that the maximum difference in sSFR between environments is $\sim$0.15 dex (for both SF and Q galaxies)}.  
We noted in that the variations of the features of the sSFR-$\sigone$ diagram with the environment decrease with $\Ms$.  Indeed the red points in \fig{trendswithqf} could just as well be fit with horizontal lines.  However, the overall pattern in \fig{trendswithqf} suggests that the weaker environmental dependence of sSFR-$\sigone$ features at high masses (\eg, \fig{ssfrvssigone}$g$ and $h$) can be attributed simply to the higher $\fq$ of all massive galaxies.
Thus, it appears that the main features of the sSFR-$\sigone$ diagram, \ie, the positions of the SF, GV and Q populations, are well predicted by the environment ($\fq$) and $\Ms$.

\section{Discussion}
\label{discussion}

\label{effects}

We have shown that the sSFR-$\sigone$ diagram is a promising diagnostic of the differences in the quenching processes among different populations of galaxies.  We have studied this diagram as a function of three population variables: stellar mass, halo mass (for satellites), and radial location within haloes (also for satellites).  Different features are seen, which may ultimately be linked to the different quenching mechanisms that are thought to affect galaxies in different environments. 

The basic result is that the sSFR-$\sigone$ diagram varies with environment: the diagram for field galaxies differs from that of all satellites, and the diagrams for satellites differ depending on group-centric distance and halo mass.  In particular, we find that:

\begin{enumerate}

 \item \jwb{{\bf ``Environment'' can be approximated by the quenched fraction.}  The environment of galaxies is in principle a function of both group-centric distance and the mass of the host halo.  We confirm what has been seen in many studies that the quenched fraction increases with $\Mh$, and in massive haloes, with decreasing $\dist$.  However, we find that these two quantities conspire such that a single quantity, the quenched fraction of the population (including in the field) accurately predicts the rest of the distribution of galaxies in the sSFR-$\sigone$ plane.  In other words, using the quenched fraction as a proxy for the environment, the following features are seen to vary smoothly with that measure:
}
 
 \item {\bf The median value of $\sigone$ in the GV decreases smoothly with the quenched fraction of the environment.}  In other words, the position of the GV bridge between the SF and Q populations in the sSFR-$\sigone$ diagram shifts  \refc{toward lower $\sigone$ by as much as 0.2 dex (at $\Ms=10^{9.75-10}$)} for satellites compared to the field, and for satellites in more massive haloes and lower cluster-centric distance.
 
 \item {\bf The sSFR ridge-line of SF galaxies falls slightly with increasing quenched fraction of the environment.}  This is a small effect ($\sim 0.1$ dex) but it is systematic over a large range of $\Ms$.
 
 \item {\bf The median $\sigone$ of quenched galaxies in a given mass bin is always higher than the star-forming population by about 0.2-0.3 dex.}  The exact value also varies slightly with environment (as measured by the quenched fraction), but the variation is smaller than the variation in the median $\sigone$ of the GV or of the sSFR of the SF population.  
 
\end{enumerate}

The magnitude of the above variations with environment depends on $\Ms$.  The quenched fraction, median $\sigone$ of the GV and sSFR of the SF population vary significantly with environment for low-mass galaxies ($\Ms \ltsima 10^{10} \Msun$) but much less for the most massive galaxies ($\Ms \gtsima 10^{10.5}\Msun$).  This may be a reflection of the uniformly high quenched fractions for intermediate- and high-mass field galaxies (\fig{trendswithqf}).  Since at a given mass the number of field galaxies is about twice the number of satellites, most quenched galaxies of mass $\Ms \gtsima 10^{10.5}\Msun$ likely quenched as field galaxies, some of which later became quenched satellites.  Galaxies of lower mass on the other hand have a quenched fraction that is more than double the quenched fraction in the field, indicating that a significant portion of these satellites quenched as satellites.  

These results are consistent with the expectation of an environmentally induced quenching track that occurs in clusters \jwb{today} and that is in addition to the quenching that occurs in the field (\fig{cartoon}).  In the field, it appears that galaxies must have high $\sigone$ relative to the rest of the SF main sequence \refc{(by about 0.2-0.3 dex)} in order to start quenching (\ie, to bring them to the GV; this is consistent with \citealp{cheung12,fang13,barro15}).  Since external influences are presumably lacking in the field, quenching must occur due to the properties of the galaxy itself or its own halo, \eg, the mass of its halo, AGN state, stability of the disc, etc.  
Referring to these collectively as ``self-quenching'', the sSFR-$\sigone$ diagram in the field indicates that high $\sigone$ is associated with self-quenching.  

In contrast, the distribution of $\sigone$ for satellites in the GV is more similar to the distribution of $\sigone$ on the SF main sequence (blue histograms in \fig{histograms}, comparing $a$ to $b$ and $g$ to $h$), indicating that quenching in clusters is not associated with high $\sigone$.  This is consistent with the influence of external factors that quench (at least to the GV) regardless of $\sigone$ or without altering $\sigone$.  Furthermore, the small suppression of the sSFR of the SF population, also independent of $\sigone$ (\fig{histograms}B and D), may be an indication that the start of quenching occurs immediately upon becoming a satellite.  Our finding that the features of the sSFR-$\sigone$ diagram vary smoothly with the environment (via $\fq$) may reflect the decreasing relative importance of self-quenching vs. external processes in more extreme enviroments.

\subsection{Why is $\sigone$ of Quenched Galaxies Always High?}

While our findings indicate that satellite quenching {\it begins} at a wide range of $\sigone$, satellites that have more or less {\it finished} their quenching all have the highest $\sigone$ \jwb{(compared to SF and GV galaxies) in a given mass bin}.  
\jwb{
As detailed in \sec{introduction}, it has been demonstrated that quenching is strongly associated with early-type morphology for the galaxy population as a whole, which is dominated by field galaxies.  However, we have shown here that even in the densest environments, there are almost no quenched satellites with low $\sigone$.  This is a surprising result in light of the various proposed satellite quenching mechanisms.  
This result rules out the simple vertically downward quenching track on the sSFR-$\sigone$ diagram, such as might be expected, for example, from the cut-off of accretion \citep{dekbir06} and the stripping of cold gas via ram pressure.  Both of these scenarios are expected to quench satellites quickly regardless of $\sigone$ (although see \citealp{mccarthy08}).  }
\jwb{
To explain high $\sigone$ in quenched satellites, we discuss here some of the remaining scenarios which can be divided into two classes: those that involve the creation of new stars in the central kpc, and those in which the $\sigone$ value of {\it individual} galaxies remains intact.  
}

Mechanisms for satellite quenching that involve the creation of new stars include galaxy-galaxy interactions (sometimes called ``harassment), which build the density in the inner kpc with each successive interaction (\citealp{lake98}).  Ram pressure, usually thought to strip gas from a galaxy, can also potentially compress gas in the inner regions. Dekel et al. (in preparation) estimate that starting at 0.5$\Rvir$, ram pressure can increase $\sigone$ by up to a factor of 2 (what is needed to explain the sSFR-$\Ms$ diagram) if the gas fraction is $\sim 0.5$.   
\jwb{Such gas fractions are seen in lower mass galaxies than studied here, or at higher $z$, and thus this process may be relevant for those regimes.
However, the galaxies in the GV observed here likely have much lower gas fractions, and so ram pressure compression may not be responsible for the high $\sigone$ of quenched satellites relative to GV satellites observed here.}

Tidal compression may prove to be more promising.  Tidal forces in haloes with shallow profiles are compressive and can increase $\sigone$ by a factor of 2 even with gas fractions as low as 0.1 (\citealp{dekel03tidalcompression}, Dekel et al., in preparation).  However, such tidal compression is only efficient within 0.1$\Rvir$.  

In any case, our results imply that any quenching scenario that increases $\sigone$ must begin quenching (bringing the galaxies from the SF to the GV) {\it before} the increase of $\sigone$.  This is indicated by the fact that the satellite quenching tracks increase in $\sigone$ only after galaxies have passed through the GV and are en route to the quenched region.

It is also possible to explain the high apparent $\sigone$ of quenched galaxies without increasing $\sigone$ in individual galaxies.  This is possible if galaxies lose mass during quenching and thus move from one mass bin to another.  For example, tidal stripping of a satellite in a higher mass bin can strip both gas and stars, potentially lowering $\Ms$ significantly from the infall value (as much as 30-40\% or more per pericentre passage - \citealp{zentner03,taylor04}), but will leave $\sigone$ intact.  Indeed, comparing the sSFR-$\sigone$ diagrams in \fig{ssfrvssigone}, $\sigone$ of the GV for satellites is always comparable to the $\sigone$ of the Q population in the immediately lower mass bin, supporting the plausibility of the stripping scenario.  The position of the GV in the sSFR-$\sigone$ diagrams would imply that the start of quenching occurs before the stripping, which is reasonable considering that tidal stripping is most efficient at the orbital pericentre.  This would be in line with the findings of \citet{pasquali10} who find that the mass-metallicity relation of satellites is offset toward lower mass.

{
One additional effect that can explain the high $\sigone$ of quenched galaxies without increasing $\sigone$ is one that has received little attention so far - namely, ``progenitor bias''.  
The idea is that 
at any given $\Ms$, Q galaxies that quench at earlier times are expected be denser on average than at later times, simply because progenitor star-forming galaxies, their dark matter halos and the whole Universe are denser at earlier epochs (e.g., \citealp{valentinuzzi10,carollo13,poggianti13,lillycarollo16}).
Thus, the sSFR-$\sigone$ diagrams for the field population can be explained by the gradual displacement of the GV bridge \refc{toward lower $\sigone$} over time, filling the \refc{lower} end of the Q cloud. The evolution of the $\sigone$ vs. $\Ms$ relation of the whole Q galaxy population averaged over all environments shows a decrease of $\sim 0.3$ dex in $\sigone$ since a redshift of $z \sim 3$ (\citealp{barro15}), which well agrees with the width of the Q cloud. 
}

As for the satellites, the morphology of the sSFR-$\sigone$ diagram could arise from the cumulative effects of $(1)$ a predominant role of field-like self-quenching at early epochs;
and $(2)$ the switching-on of satellite-quenching at later epochs, which quenches at lower $\sigone$. 
(See also \citealp{gallart15} who suggest that dwarf satellite morphologies are determined at birth.)

{It is possible that the $\sigone$-increasing mechanisms (such as those described above) and the effect of progenitor bias work in concert.  How important these effects are can at least in part be tested through direct measurement of sSFR-$\sigone$ diagrams at earlier epochs, and their stellar populations (work in progress).  
}

{\bf In summary}, we have used the sSFR-$\sigone$ diagram as a useful diagnostic of quenching among different populations of galaxies. 
 The power of $\sigone$ to predict quenching that has been seen
     in bulk populations of galaxies is confirmed when galaxies
     are divided up into field vs satellites and satellites
     in different haloes and at different group-centric distances.    
  The location of the transitioning galaxies (\ie, the GV) in the sSFR-$\sigone$ diagram varies smoothly with the
    environment, $\sigone$ being lower for satellites in larger halos and at
    smaller radial distances within the same-mass halos.  
    We interpret this shift as indicating the relative importance in different environments of the field quenching track vs. the cluster quenching track.  In all environments, the Q population always has high $\sigone$ 
    relative to their masses.   We proposed two types of scenarios to explain the high $\sigone$ of Q satellites relative to the low $\sigone$ of GV (and SF) satellites.  One class of scenarios involves the creation of new stars in the inner kpc \jwb{(harassment, ram pressure compression and tidal compression)} while the other class does not involve such star formation \jwb{(tidal stripping causing significant mass loss and progenitor bias)}.  
     It is the goal of future work to distinguish between these possibilities.

\section*{Acknowledgements}
We acknowledge the
helpful and stimulating discussions with 
Guillermo Barro,
Colin DeGraf,
Marla Geha,
Will Hartley,
Andrew Hearin,
David Koo,
Simon Lilly,
Nir Mandelker,
 Joel Primack, 
 Aldo Rodgriquex-Puebla,
 Kevin Schawinski,
Rachel Sommerville,
Benny Trakhtenbrot,
and 
Frank van den Bosch.
We acknowledge support from the Swiss National Science Foundation. SF acknowledges support from NSF grant AST-08-08133.

Funding for the SDSS and SDSS-II has been provided by the Alfred P. Sloan Foundation, the Participating Institutions, the National Science Foundation, the U.S. Department of Energy, the National Aeronautics and Space Administration, the Japanese Monbukagakusho, the Max Planck Society, and the Higher Education Funding Council for England. The SDSS Web Site is http://www.sdss.org/. The SDSS is managed by the Astrophysical Research Consortium for the Participating Institutions. The Participating Institutions are the American Museum of Natural History, Astrophysical Institute Potsdam, University of Basel, University of Cambridge, Case Western Reserve University, University of Chicago, Drexel University, Fermilab, the Institute for Advanced Study, the Japan Participation Group, Johns Hopkins University, the Joint Institute for Nuclear Astrophysics, the Kavli Institute for Particle Astrophysics and Cosmology, the Korean Scientist Group, the Chinese Academy of Sciences (LAMOST), Los Alamos National Laboratory, the Max-Planck-Institute for Astronomy (MPIA), the Max-Planck-Institute for Astrophysics (MPA), New Mexico State University, Ohio State University, University of Pittsburgh, University of Portsmouth, Princeton University, the United States Naval Observatory, and the University of Washington.

\bibliographystyle{mnras}
\bibliography{jobib}

 \renewcommand{\theequation}{A\arabic{equation}}
  \renewcommand{\thesection}{A\arabic{section}}
  \renewcommand{\thefigure}{A\arabic{figure}}
    \setcounter{equation}{0}    \setcounter{section}{0}    \setcounter{figure}{0}  
  \section*{APPENDIX }
  \label{appendix_dots}

\begin{figure*}
\includegraphics[width=0.7\textwidth]{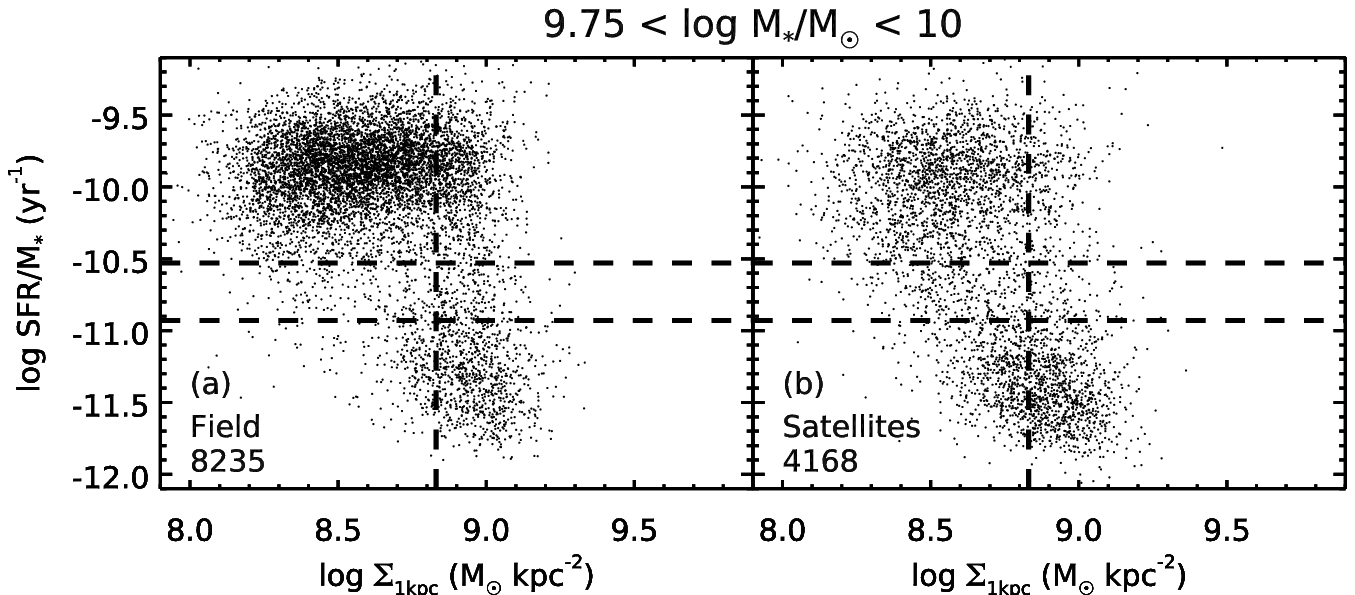}
\includegraphics[width=0.7\textwidth]{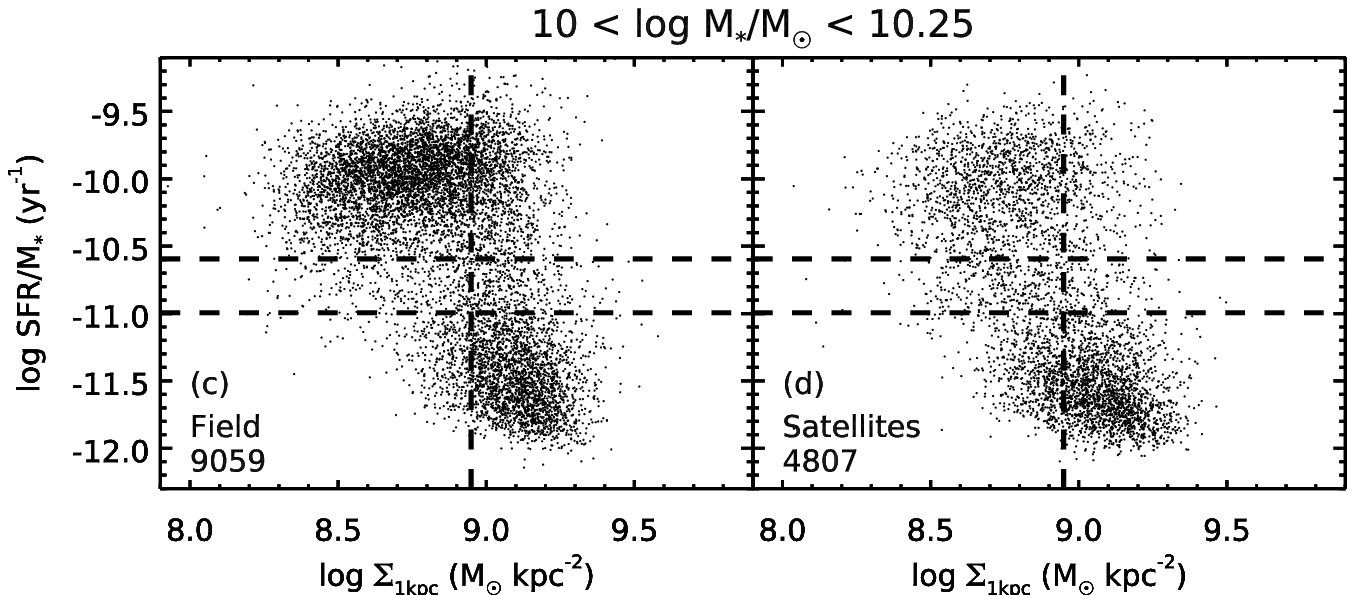}
\includegraphics[width=0.7\textwidth]{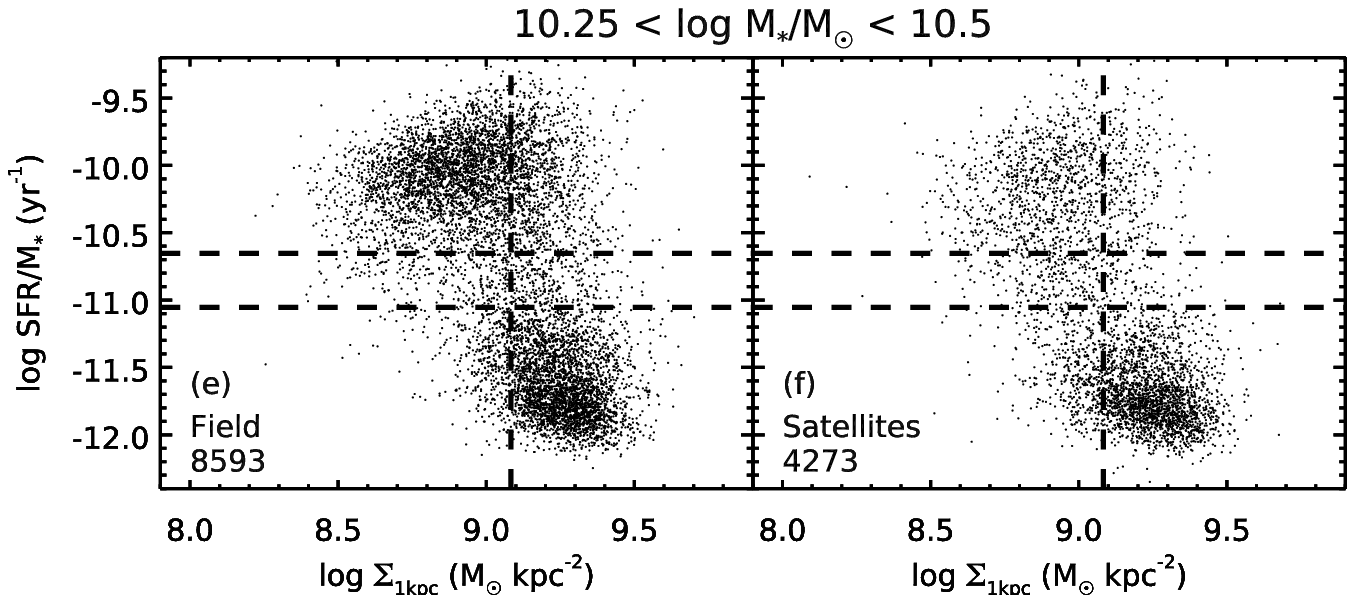}
\includegraphics[width=0.7\textwidth]{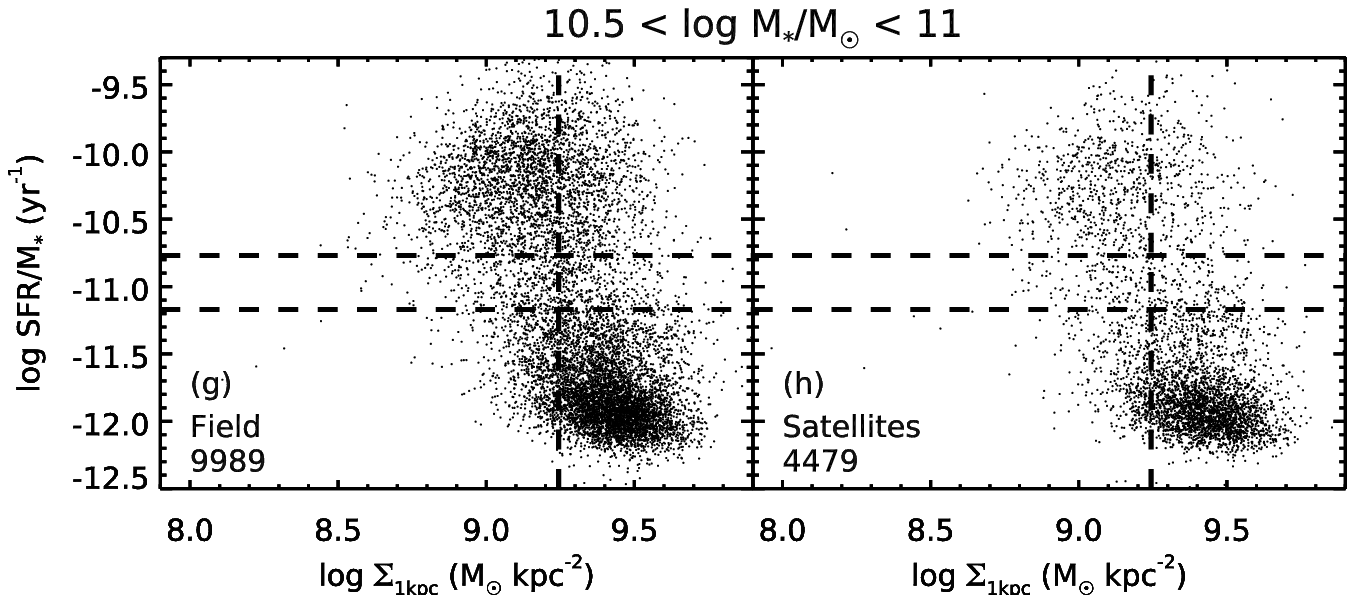}
\caption{\small sSFR vs. $\sigone$, in four bins of $\Ms$ for field galaxies and 
  satellites.  This is the same as \fig{ssfrvssigone} but using the raw unweighted and unsmoothed points.  
  The GV as defined in \fig{gmsdefs} lies rougly between the two horizontal lines.  
  For reference, the vertical line marks the median $\sigone$ of hte field GV.
  The sample size of each panel is indicated in the bottom corner.
   }
  \label{ssfrvssigone_dots}
\end{figure*}

\begin{figure*}
 \includegraphics[width=0.8\textwidth]{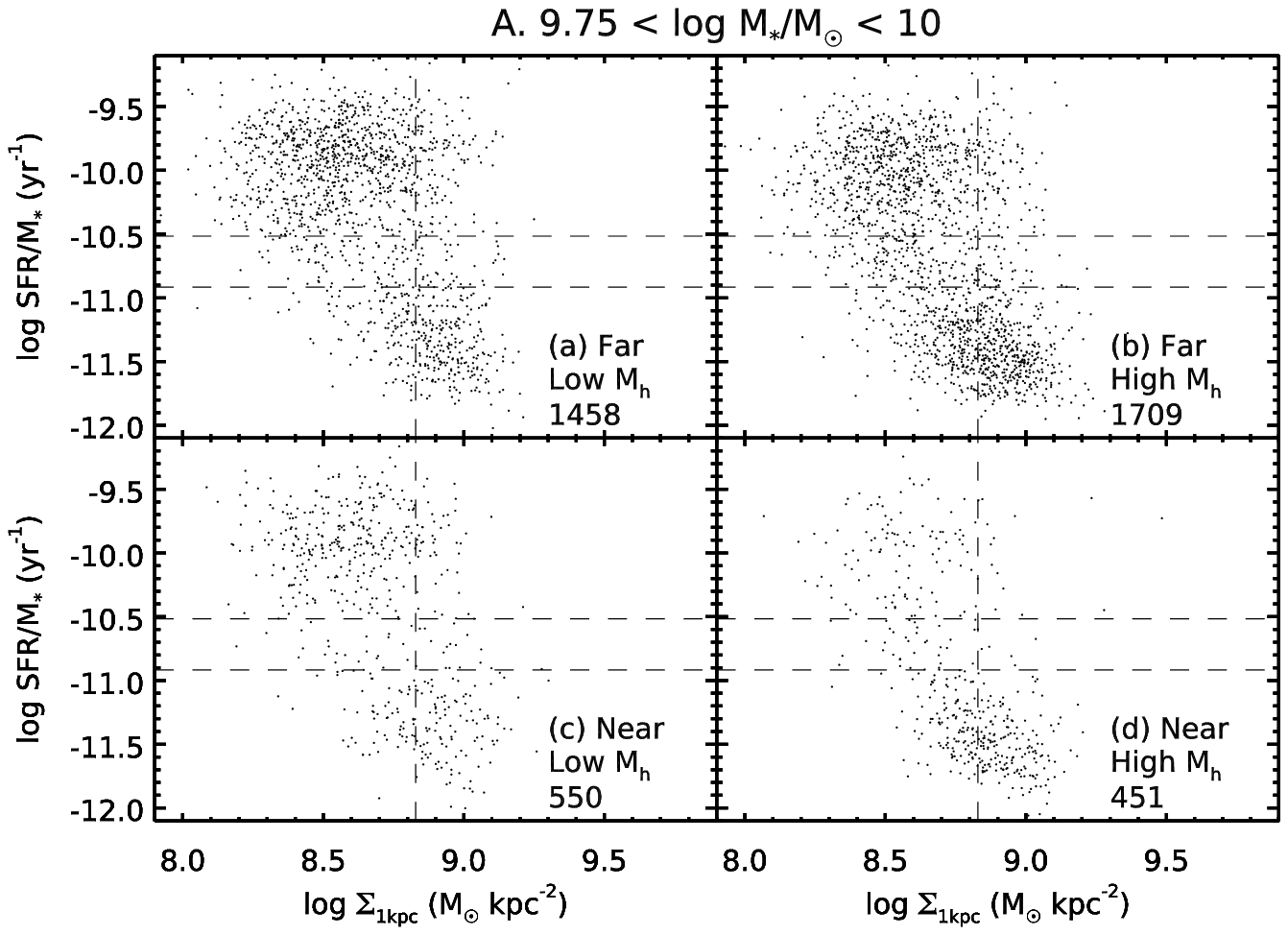}
  \includegraphics[width=0.8\textwidth]{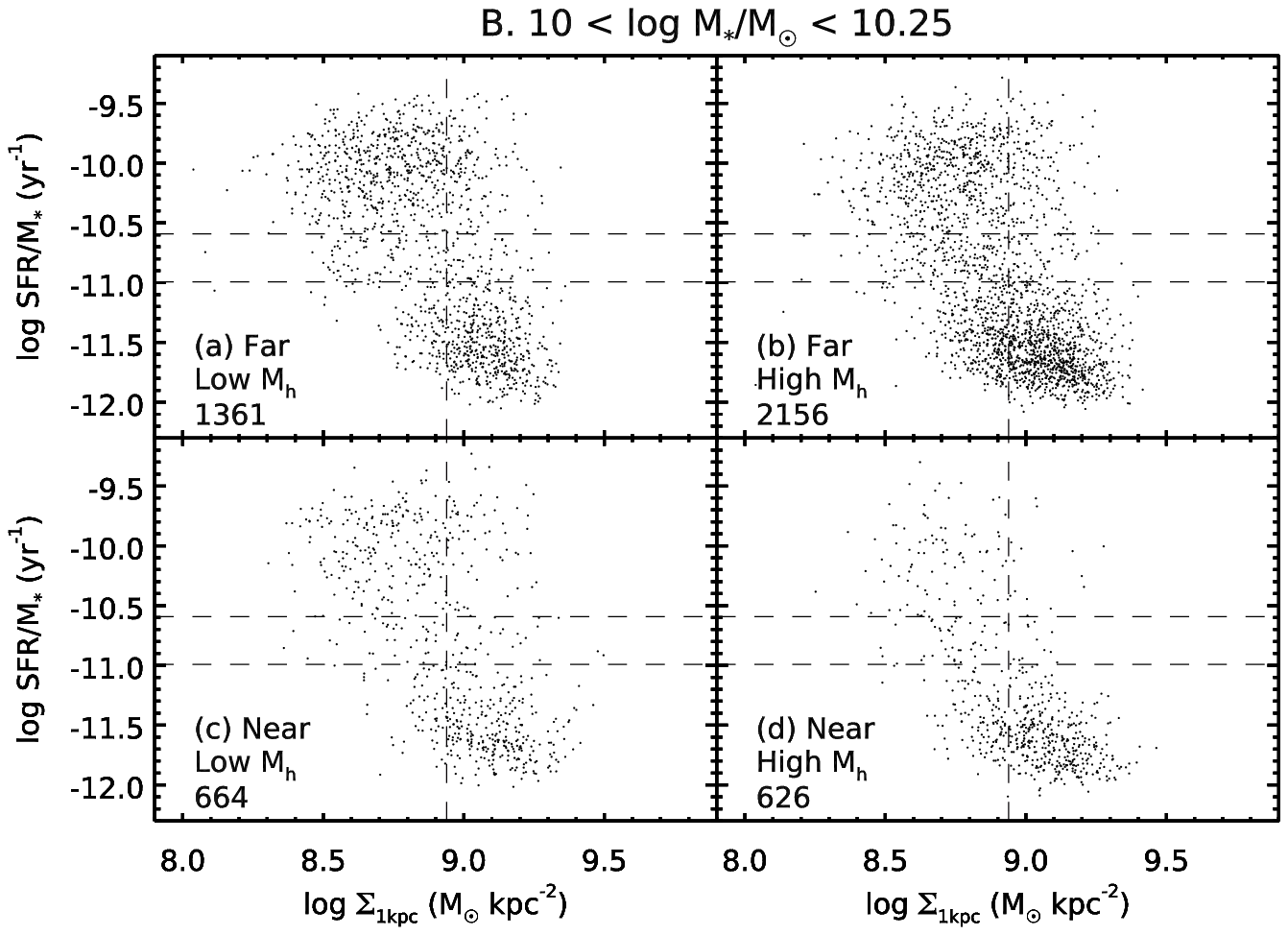}
 \caption{The distribution of satellite galaxies in the sSFR-$\sigone$ plane in four panels representing different environments in two mass bins (A and B).  This is the same as \fig{panelsdvsmh_lowm} but using the raw unweighted and unsmoothed points.
 ``Near'' and ``Far'' refer to log $\Dist$ less than or greater than -0.6 (or $\Rproj \sim 0.25\Rvir$).  ``Low $\Mh$'' and ``High $\Mh$'' refer to haloes less massive than or more massive than $10^{13.5}\Msun$. 
 Satellites in the GV as defined in \fig{gmsdefs} lie roughly between the two horizontal lines.  For reference, the vertical line marks the median $\sigone$ of the field GV.
  The sample size of each panel is indicated in the bottom corner.   
  }
 \label{panelsdvsmh_lowm_dots}
\end{figure*}

\begin{figure*}
 \includegraphics[width=0.8\textwidth]{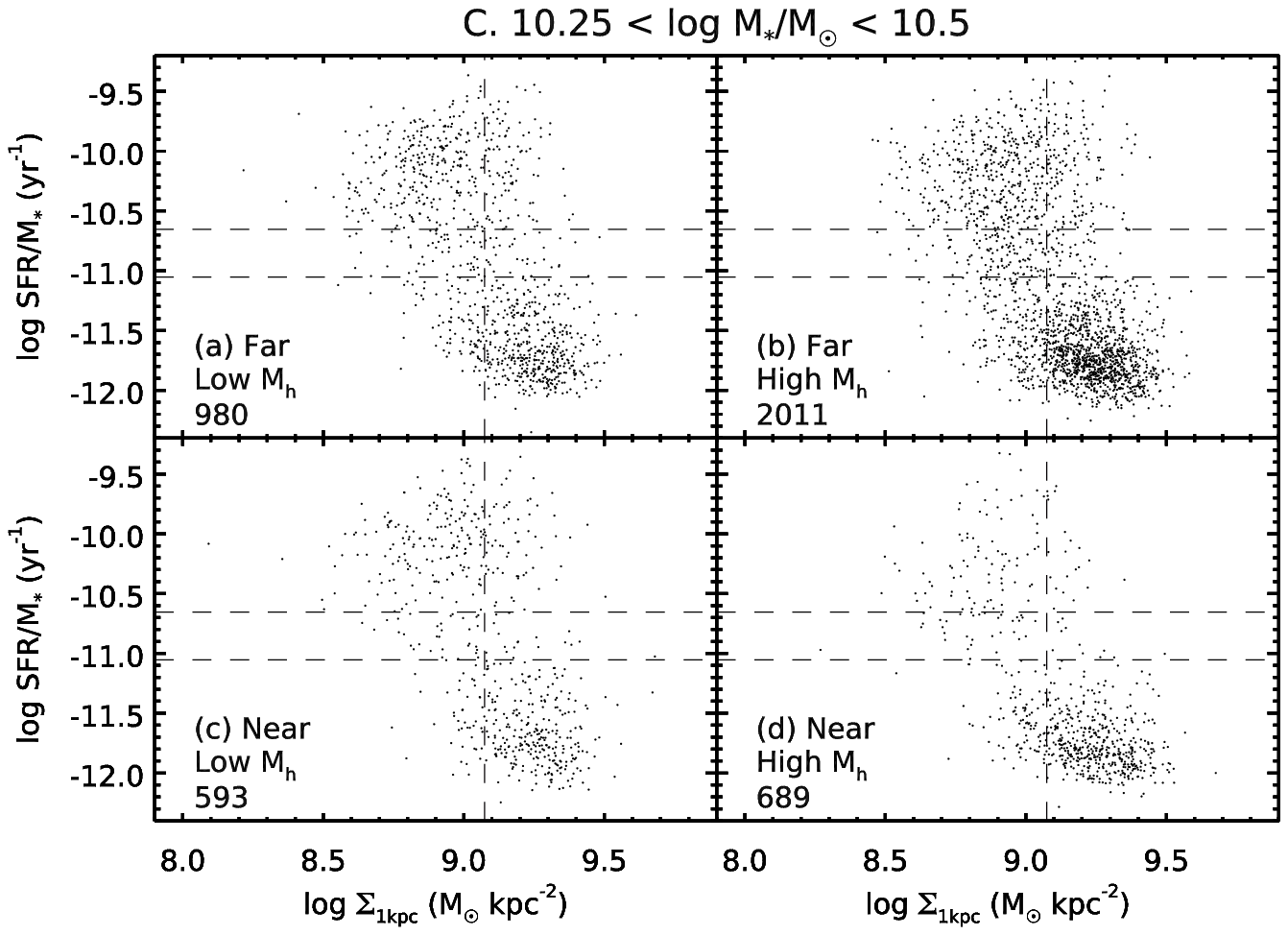}
  \includegraphics[width=0.8\textwidth]{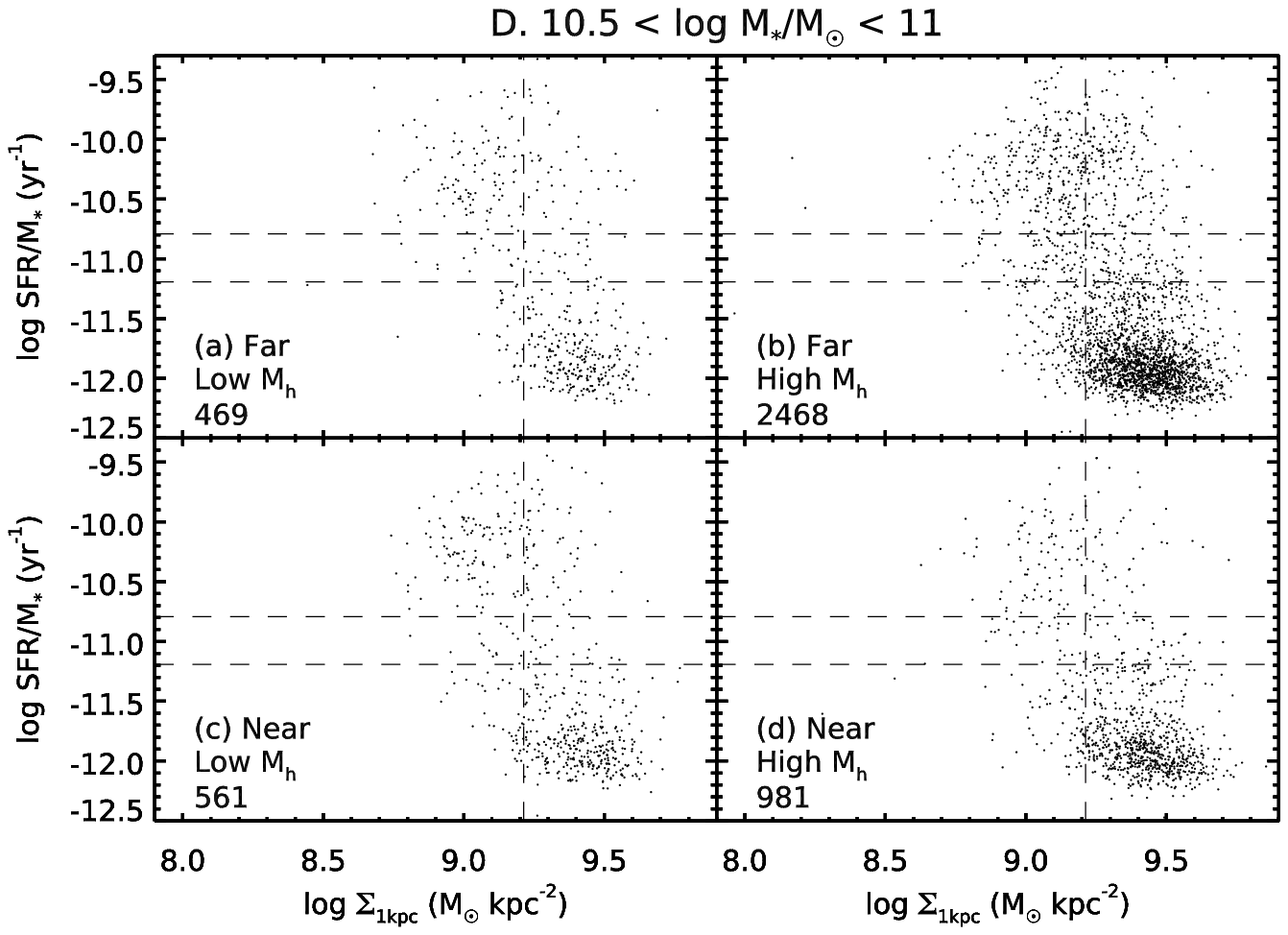}
\caption{
The distribution of satellite galaxies in the sSFR-$\sigone$ plane in four panels representing different environments in two mass bins (C and D).  This is the same as \fig{panelsdvsmh_highm} but using the raw unweighted and unsmoothed points.
 ``Near'' and ``Far'' refer to log $\Dist$ less than or greater than -0.6 (or $\Rproj \sim 0.25\Rvir$).  ``Low $\Mh$'' and ``High $\Mh$'' refer to haloes less massive than or more massive than $10^{13.5}\Msun$. 
 Satellites in the GV as defined in \fig{gmsdefs} lie roughly between the two horizontal lines.  For reference, the vertical line marks the median $\sigone$ of the field GV.
  The sample size of each panel is indicated in the bottom corner.   
   }
 \label{panelsdvsmh_highm_dots}
\end{figure*}

Fig.s \ref{ssfrvssigone_dots} to \ref{panelsdvsmh_highm_dots} are the same as Fig.s \ref{ssfrvssigone}, \ref{panelsdvsmh_lowm} and \ref{panelsdvsmh_highm}, except that they show individual points instead of weighted contours.

\label{lastpage}

\end{document}